# Techno-economic analysis of solar PV power-to-heat-to-power storage and trigeneration in the residential sector

A. Datas[1,2], A. Ramos[1,2], and C. del Cañizo[1]

[1]Instituto de Energía Solar, Universidad Politécnica de Madrid, 28040 Madrid, Spain.

[2]Universitat Politècnica de Catalunya, Jordi Girona 1-3, Barcelona 08034, Spain.

**Keywords**: PHPS, CCHP, TES, trigeneration, cogeneration, energy storage, heat, thermal, photovoltaics, solar, self-consumption, thermophotovoltaics, Stirling, ultra-high temperature.

## Abstract

This article assesses whether it is profitable to store solar PV electricity in the form of heat and convert it back to electricity on demand. The impact of a number of technical and economic parameters on the profitability of a self-consumption residential system located in Madrid is assessed. The proposed solution comprises two kinds of heat stores: a low- or medium-grade heat store for domestic hot water and space heating, and a high-grade heat store for combined heat and power generation. Two cases are considered where the energy that is wasted during the conversion of heat into electricity is employed to satisfy either the heating demand, or both heating and cooling demands by using a thermally-driven heat pump. We compare these solutions against a reference case that relies on the consumption of grid electricity and natural gas and uses an electrically-driven heat pump for cooling. The results show that, under relatively favourable conditions, the proposed solution that uses an electrically-driven heat pump could provide electricity savings in the range of 70 – 90% with a payback period of 12 – 15 years, plus an additional 10 – 20% reduction in the fuel consumption. Shorter payback periods, lower than 10 years, could be attained by using a highly efficient thermally driven heat pump, at the expense of increasing the fuel consumption and the greenhouse gas emissions. Hybridising this solution with solar thermal heating could enable significant savings on the global emissions, whilst keeping a high amount of savings in grid electricity (> 70 %) and a reasonably short payback period (< 12 years).

1. **Introduction**

The energy consumption in buildings represents around 40% of the world energy consumption and one-third of global $CO_2$ emissions [1]. Thus, during the last decade a number of technological solutions have been proposed targeting the improvement of the energy efficiency and the reduction of the $CO_2$ emissions associated to energy consumption in the building sector. Among them, the so-called combined cooling, heating and power (CCHP), or trigeneration systems [2]–[6] are of especial interest, as they are intended to reduce the emissions associated to the three main kinds of energy demands in a building: electricity, heating and cooling. A typical CCHP system comprises a power generation unit (PGU), e.g. a microturbine, and a thermally-driven heat pump (THP) [7], e.g. an absorption chiller. The PGU is typically powered by natural gas and produces electricity and heat as by-product. This heat can be directly used for heating in winter or transferred to a THP for cooling in summer. Compared with the solution based on large centralized power plants and local air-conditioning systems, distributed CCHP provides a significant improvement of fuel utilization, in the range of 70 to 90% with respect to 30 – 45% of centralized power plants [2]. Further reduction of the emissions is possible by fully or partially replacing natural gas by solar thermal heating [8]. Most of the studies consider organic Rankine cycles (ORC) [9], [10] in this case, as they enable operation temperatures below 220 ºC, meeting the requirements of non-concentrating solar collectors that could be used in residential and commercial environments. However, the low exergy content of such a low-grade heat results in very low PGU conversion efficiencies, typically below 10 – 15% [9], [11], [12], which precludes the achievement of a clear economic advantage.

In the last decade, the dramatic cost reductions of solar PV technology have triggered the interest on self-consumption of PV electricity in both commercial and residential buildings [13]. With an average growth rate of 51% [14], new solar PV power additions in 2017 accounted up to 98 GW, out-stripping the 70 GW of net fossil fuel generating capacity added the same year. Small-scale solar PV systems in distributed applications, mostly building integration, accounts for a significant share of these new PV additions (~ 38% [15]). Only in China, distributed capacity additions in 2017 summed up to 19.4 GW, and new rooftop systems saw a three-fold increase relative to 2016 [16]. Solar PV installations are simple, reliable, and do not require high maintenance. Thus, they are very appealing to produce electricity in small distributed applications. In the residential context, many solutions have been proposed to integrate solar PV systems with CCHP systems, including the hybridization of solar PV with gas-powered CCHP [17]–[20], the use of hybrid PV/thermal (PVT) solar collectors, using both non-concentrating [10], [21]–[24] and concentrating technologies [25], the integration of PV systems with electrically-driven heat pumps (EHP) [26]–[28], as well as the direct use of PV electricity for heat production [29].

In all these cases, the self-consumption of PV electricity is limited due to the lack of synchronization between solar irradiance and local consumption. For systems where PV electricity is used exclusively to satisfy the electric power demand, self-consumption ratio (defined here as the fraction of the PV generated electricity that is used to supply the loads) is usually limited to 25 – 30% [30]. Self-consumption can be increased in systems where solar PV electricity is also used to provide heating and cooling [13], [26], [31], [32]. However, in this case, the attainable self-consumption ratio strongly depends on the climate conditions and the building standardized heating load. For instance, in old building standards located in Stuttgart (Germany), the combination of a heat pump and a hot water storage enable self-consumption increments of ~ 35% [26]. However, for next generation zero-energy buildings the same solution provides self-consumption potential increments of only ~ 10%, as energy demand of appliances dominates compared to the electrical energy demand to drive the heat pump [26].

To increase the self-consumption levels, solutions must integrate a system that stores the excess of PV electricity, such as electrochemical batteries [17]–[19], [33]. However, current prices for stationary residential battery storage are prohibitive, exceeding 1000 US$$_{2015}$ per kWh of electricity storage capacity [34]. Even in the case of reaching global cumulative storage capacities of 1000 GWh (from a current cumulative capacity of ~ 1 GWh), empirical learning curves project that the future cost of stationary residential electricity storage, regardless of the technology type, will be in the range of ~ 340 $$_{2015}$/kWh$_{el}$ [34]. Additionally, in most CCHP systems lifetime (> 20 years), the batteries require replacement (4 -15 years) and this can have a significant impact on the lifetime cost of the full installation [17]. Consequently, when looking at the profitability of the CCHP solution, the high capital cost of batteries results in optimal systems with a relatively small storage capacity, consequently providing small self-consumption improvements in the range of 13 – 24% [13]. For this reason, finding low-cost alternatives for electricity storage in the residential sector is an important field of research today [35], [36].

Although it might sound counterintuitive, among the many possible energy storage options, those that are particularly interesting for CCHP applications are those with low round-trip (electric-to-electric) efficiency, as they may deliver part of the stored energy in a form of heat. Some examples are compressed air [37] or hydrogen storage combined with fuel cells [38]. In these cases, the comparatively low round-trip efficiency can be compensated by the (eventual) lower cost of the technology and the profitable use of the exhaust heat. In this regard, a potentially low-cost alternative for electricity storage that has not received much attention is the power-to-heat-to-power storage (PHPS) concept. PHPS involves the conversion of electricity into heat, which is then stored and later converted back to electricity on demand. The dramatic cost reduction of solar PV electricity along with the potentially lower capital costs of PHPS might result in a profitable PV+PHPS solution in the context of CCHP applications, where the low-grade heat produced during the heat-to-power conversion process may be used for satisfying both heating and cooling demands. Today, the use of thermal storage for power generation is virtually limited to concentrated solar power (CSP) plants, where the store temperature rarely surpasses ~ 500 ºC [39]. On the contrary PHPS systems could theoretically reach ultra-high temperatures (> 1000 ºC), subsequently enabling very high heat-to-power (H2P) conversion efficiencies of ~ 40% or beyond. This is around 3 - 4 times greater than that of low-temperature solar thermal ORC systems discussed above [9], [11], [12]. A hypothetical system comprising a 40% efficient solar thermal collector and a 10% efficient ORC [40], [9] would produce an overall solar-to-electric conversion efficiency of 4%, which is half the efficiency than that of a PHPS system comprising 20% efficient solar PV modules and 40% efficient PGU. Besides, the use of high grade heat store would potentially enable much more efficient THP cooling, as well as higher stored energy densities [41], [42], which is a remarkable advantage in space constrained residential applications.

Recent studies have established a few conceptual PHPS embodiments, which differentiate in the way that heat is produced, stored, and converted back into electricity. A particularly promising concept is pumped heat electricity storage (PHES), in which a high temperature heat pump cycle transforms electricity into heat, which is stored inside two large regenerators, and a thermal engine cycle transforms the stored heat back into electricity [43]–[45]. PHES has a high theoretical round-trip efficiency (RTE) in the range of 40 – 70%, depending on the operational temperature range [44]. Predominately conceived for large grid-electricity storage applications, the potential viability of PHES in the residential sector has not been assessed yet. Other conceptually simpler approaches consider the use of ultra-high temperature (> 1000 ºC) joule heating for sensible- [46] and latent- [41] heat storage combined with a thermophotovoltaic (TPV) power generation. Despite having lower RTE potential (less than ~ 40%), these designs might bring some advantages, as the modularity and the lack of moving parts. Solutions based on the use of high temperature heat pumps have been also recently proposed to mitigate the high thermal losses that could be eventually derived from operating at ultra-high temperatures; theoretically enabling an increment of the round-trip conversion efficiency up to ~ 50% [47], [48].

Regardless of the particular system implementation, it is still unclear under what circumstances a PHPS system could be profitable. In [46], [49] the minimum tolerable RTE of an energy storage system used for grid-electricity storage is estimated at 36% for the case of Pennsylvania-New-Jersey-Maryland grid in 2017. However, this study assumes that the electricity is purchased from the grid, and therefore, the minimum RTE, is entirely determined by the ratio between on-peak and off-peak electrical prices in a very specific case. Besides, the exhaust heat produced by the PGU is not used, which is detrimental for the profitability of a PHPS solution.

In this study we assess the integration of a PHPS system in a CCHP solution for the self-consumption of solar PV electricity in the residential environment. To the best of our knowledge, no comprehensive techno-economical assessment of PV generation coupled with a PHPS system has been evaluated so far. Thus, we will answer some of the most fundamental questions regarding the profitability of this solution, such as the maximum cost, the minimum PGU conversion efficiency, or the maximum heat insulation losses that are tolerable in order to provide reasonably low payback periods and significant energy savings.

## 2. System model and methodology

*Figure 1* shows the three kinds of system configurations that are analysed in this work. The reference case (*Figure 1*-a) comprises a conventional boiler (for heating) and an EHP (for cooling). In this case, all the energy consumption, either electricity ($C_e$) or heat ($C_h$), is obtained from the retail markets. This reference case will be used to evaluate the relative improvements of the proposed solutions incorporating a PHPS system. *Figure 1*-b illustrates a PHPS configuration comprising a solar PV system, high-grade thermal energy store (HTES), a PGU, a low- or mid-grade thermal energy store (LTES), and an EHP. Grid electricity and fuel are used as backups to ensure reliability of energy supply. In what follows, this configuration will be named PHPS-E. *Figure 1*-c illustrates a very similar configuration but using a thermally-driven heat pump (THP), instead of an EHP. This configuration will be named PHPS-T. In both configurations the PV electricity ($G_e$) is either i) used directly to satisfy the electric demand, ii) stored as high-grade heat in the HTES ($P_{in,HTES}$), iii) stored as low-grade heat in the LTES ($P_{in,LTES}$), or iv) lost ($P_{loss,PV}$) if there is no electricity consumption and both HTES and LTES stores are at their maximum capacities ($E_{HTES,max}$ and $E_{LTES,max}$, respectively). The possibility of selling the excesses of PV electricity to the grid is not considered in this study. The high-grade heat stored in the HTES can be supplied on demand to the PGU ($Q_{in,PGU}$) to produce electricity with a conversion efficiency $\eta_{PGU}$ ($P_{out,PGU} = \eta_{PGU} Q_{in,PGU}$). During this process, the produced low-grade exhaust heat $Q_{out,PGU} = (1 - \eta_{PGU})Q_{in,PGU}$ can be either stored in the LTES ($Q_{in,LTES}$), or lost ($Q_{loss,PGU}$) if the LTES is at maximum capacity. Additional losses take place due to the non-ideal thermal insulation of the LTES ($Q_{loss,LTES}$) and the HTES ($Q_{loss,HTES}$). All these kinds of losses ($Q_{loss} = Q_{loss,HTES} + Q_{loss,LTES} + Q_{loss,PGU} + P_{PV,loss}$) contribute to reduce the self-consumption ratio, defined in this work as $SC = 1 - \sum_{t=1}^{8760T} Q_{loss}(t) / \sum_{t=1}^{8760T} G_e(t)$, being $8760 \times T$ the total number of hours during the entire system lifetime $T$ (in years).

The cooling power ($C_c$) is satisfied by either an EHP (case PHPS-E, *Figure 1*-b) or a THP (case PHPS-T, *Figure 1*-c). The former case implies an extra consumption of electricity ($P_{EHP} = $ $C_c/COP_{EHP}$), whereas the latter implies an extra consumption of heat ($C_{hc} = C_c/COP_{THP}$). The coefficient of performance (COP) of each device indicates the ratio between the useful cooling provided to the input energy, either heat (THP) or electricity (EHP). The low-grade heat in the LTES can be used either to satisfy the space heating and domestic hot water demands ($C_h = C_{hh}$) in the PHPS-E configuration (*Figure 1*-b) or both heating and cooling demands ($C_h = C_{hh} + C_{hc}$) in the PHPS-T configuration (*Figure 1*-c). In both cases, additional heating from the external boiler ($Q_{ext}$) might be necessary to ensure supply reliability. Notice that in the PHPS-T configuration (*Figure 1*-c) an additional heat coming from the boiler may be needed for satisfying the cooling demand. Hybrid absorption-compression heat pumps [50] enabling both heat and electricity inputs might be interesting in this application, but they are not considered in this study for the sake of simplicity.

In this study we use simplified models for each device of the PHPS system. The detailed model equations are shown in *Figure 2*, which also illustrates the energy management algorithm. At every time step ($\Delta t = 1$ hour), the energy rates (in kW$_{el}$ and kW$_{th}$) shown in *Figure 1*, and the stored energy (in kWh$_{th}$) in the HTES and LTES ($E_{HTES}$ and $E_{LTES}$, respectively) are calculated following the procedure illustrated in *Figure 2*. This algorithm first evaluates whether there is an excess or defect of generated PV electricity, i.e. whether the net consumed electrical power $P_{net} = C_e - G_e$ is negative or positive, respectively. If demand exceeds the PV generation ($C_e < G_e$), all the PV electricity is directly used to satisfy that demand. The additional electricity ($P_{net}$) is supplied by either the PGU, if the HTES has enough stored heat ($E_{HTES} > P_{net}\Delta t/\eta_{PGU}$), or by the electrical grid, in the opposite case. If the electrical power demand is higher than the maximum power capacity of the PGU ($P_{net} > P_{max,PGU}$), both the PGU and electrical grid contribute to satisfy such demand. The low-grade heat produced during the operation of the PGU ($Q_{out,PGU}$) may be stored in the LTES, if this is not already at its maximum capacity. In the case the PV generation exceeds the consumption of electricity ($G_e > C_e$), such excess may be stored as high-grade heat in the HTES or as low-grade heat in the LTES. In principle, the system will prioritize the charge of the HTES, as it stores a higher-grade heat that can be later converted into electricity by the PGU. There is only one scenario where this excess of PV electricity is stored in the LTES instead of HTES. This is the case that there is heat consumption ($C_h = C_{hh} + C_{hc} > 0$), the HTES charge is high ($E_{HTES} > xE_{HTES,max}$), and the LTES charge is low ($E_{LTES} < yE_{LTES,max}$). In this study we have fixed $x = 0.99$ and $y = 0.1$, which means that this situation is very improbable, and the vast majority of the excesses of PV electricity is stored in the HTES rather than in the LTES. The optimization of the values of the parameters $x$ and $y$ is out of the scope of this work. In the case that both LTES and HTES are at their maximum capacities, the PV electricity is inevitably lost, contributing to the increase of the total energy losses of the system ($Q_{loss}$). Finally, the total heat consumption ($C_h = C_{hh} + $ $C_{hc}$) is satisfied either by the LTES, the boiler, or both combined, depending on whether the amount of stored heat in the LTES is enough to fully satisfy the heat demand. The amount of heat stored in the HTES and LTES is updated at every time step by evaluating the corresponding energy balance equations, as shown at the bottom of *Figure 1*.

The model described above assumes no exergy degradation of stored heat in both HTES and LTES. A non-degraded heat flowing out of the high-grade HTES enables assuming a constant PGU conversion efficiency during the full discharge cycle. This assumption is reasonable especially for thermal stores based on latent heat, where exergy losses are minimum. The energy losses in the LTES are approximated using the standard EN 12977-3 for hot water stores, according to which the heat loss rate in W/K is $UA_{LTES} = a_{LTES}\sqrt{V_{LTES}}$, being $V_{LTES}$ the water volume in litres, and $a_{LTES}$ a constant that usually takes a value of ~ 0.1 for good insulated hot water storage stores. The overall heat losses (in W) are approximated by $Q_{loss,LTES} = a_{LTES}\sqrt{V_{LTES}}\Delta T_{LTES}$, being $\Delta T_{LTES}$ the mean difference between LTES and ambient temperatures (assumed $\Delta T_{LTES} = 70$ ºC for the PHPS-E system and $\Delta T_{LTES} = 130$ ºC for the PHPS-T one, as the later requires higher temperature to operate a highly efficient thermally driven heat pump). The heat losses in the HTES are calculated using a similar approach, so that $Q_{loss,HTES} = a_{HTES}\sqrt{V_{HTES}}\Delta T_{HTES}$, being $a_{HTES}$ an unknown parameter that should be eventually determined for high temperature heat stores. To relate the volume ($V_{HTES}$ and $V_{LTES}$) with the storage maximum capacity ($E_{HTES,max}$ and $E_{LTES,max}$) we assume an energy storage density of 0.08 kWh$_{th}$/l for the LTES, corresponding to that of a hot water store, while for the HTES, the energy density strongly depends on the storage media and its operational temperature. In this study we assume a latent heat store with an energy density that depends on the operational temperature according to the following equation: $E_{d,TES} = 6.74 \times 10^{-7}(T_{TES})^2 - 1.72 \times 10^{-4}T_{TES} + 0.140$ that has been obtained from fitting the latent heat of fusion of a few pure metals with different melting points in the range of 230 – 2070 ºC (Sn, Zn, Al, Si and B) [41]. This equation captures the fact that higher temperatures enable higher stored energy densities. For instance, a 20 kWh$_{th}$ HTES operating at 1414 ºC (which will be the case if silicon is used as phase-change material) would have an energy density of 1.24 kWh$_{th}$/l (silicon latent heat). Assuming an optimistic value for $a_{HTES}$ of 0.1, such HTES system would lose about 2.8% of its maximum stored energy in one hour. As a reference, a hot-water store of the same capacity and with $a_{LTES} = 0.1$ would lose only 0.24% per hour. For a HTES at 800 ºC, heat losses would be only very slightly reduced to 2.7% (assuming the same value of $a_{HTES}$). This is because the lower temperature also brings a lower energy density, counteracting to keep a similar amount of heat losses. The evident way to reduce the losses is the use of advanced thermal insulation designs with a very small $a_{HTES}$ value. Possible options include the use of vacuum insulation or other advance concepts, such as the use of heat pumps [47]. Whether such low $a_{HTES}$ values are attainable by practical thermal insulation systems is out of the scope of this work. Our aim is to provide its bounds in order to reach profitability in a PHPS application. An additional possibility to reduce heat losses in the HTES is to use eutectic alloys, instead of pure elements, with enhanced latent heats at lower melting temperatures, such as Al-12Si (549 kWh/m$^3$ at 577 ºC) [51] or Fe-26.3Si-9.3B (~ 1240 kWh/m$^3$ at ~ 1200 ºC) [42].

Time-dependent profiles of heating (space and hot water), cooling, and electricity consumptions ($C_{hh}$ and $C_c$, $C_e$) are obtained by the Energy Plus software [52] for a detached household with two floors with an area of 60 m$^2$ each, 30% of openings (glazing), the U-value of the façade is 0.26 W/(m$^2$·K) and the U-value of the roof is 0.18 W/(m$^2$·K). These U-values match with the guidelines provided in the *Spanish Building Code* [53], which in turn are also in agreement with the corresponding guidelines of most of the European countries [54], [55]. For the energy simulations, typical occupancy profiles of a 4-inhabitant house (2 adults and 2 children) are considered. The simulations differentiate between working and non-working days, and provide loads, schedules for lighting and home appliances, as well as for occupancy, and the air renovations in the different months. The primary and secondary set-points for air conditioning in summer (i.e. from June to September) are 25 ºC and 27 °C, respectively, while primary and secondary temperature set-points for space heating (i.e. from January to October) are 20 °C and 17 °C. DHW inlet/delivery temperature is 15/60 ºC and the normalized lighting power density in W/m$^2$ -100 lux is 5. The Energy Plus software calculates the energy consumption (for heating, cooling and DHW) of a given household for a user-defined HVAC and DHW systems, thus resultant energy consumption will depend on the conversion efficiency of the selected systems. In the particular case of the electricity, electricity demand will be equal to electricity consumption if no losses are considered. Therefore, in order to obtain energy demand profiles which will be later inputs in our system model, conversion efficiency of the HVAC and DHW systems is set to 1. Energy simulations of the reference building consider 1-hour step weather data from Energy Plus$^{TM}$ database, which includes: direct and diffuse solar irradiance, solar height and azimuth, atmospheric pressure, ambient temperature and wind direction and velocity. Finally, from the energy simulations, hourly global energy demand of the reference building over a year is calculated. Thus, 1-hour step data of electricity for lighting and other appliances, cooling, domestic hot water (DHW) and space heating demands is obtained. The hourly PV electrical power generation per kW of installed PV capacity ($G_e^*$, in equivalent hours) in Madrid is calculated by means of the PVsyst software [56]; and considering the same weather data from Energy Plus$^{TM}$ database for coherence with the time-dependent energy consumption profiles. For this calculation monocrystalline silicon PV modules of 318 Wp from the manufacturer SunPower and a Sunny Boy series inverter of the manufacturer SMA have been considered. PV modules are orientated to the South and their tilt angle is 34º, which is the optimum value that maximises the annual PV generation in Madrid. For the comprehensive system model, energy output after the inverter is considered. The total PV electricity generated during one hour ($G_e$, in kWh) is estimated from these simulations as $G_e = G_e^* P_{nom,PV}$, being $P_{nom,PV}$ the nominal PV installed power (in kW).

*Figure 3*-a shows an example of the resultant electrical (top) and thermal (bottom) energy rates along with the amount of energy stored in the HTES (top, solid line) and LTES (bottom, solid line) for a system with $\eta_{PGU} = 30\ \%$, $P_{nom,PV} = 10.6$ kW$_{el}$, $P_{max,PGU} = 1.35$ kW$_{el}$, and $E_{max,HTES} = 33.7$ kWh$_{th}$. As it will be seen in the discussion section, the sizing of the three key elements of this system, i.e. the PV, HTES and PGU, has been optimized to minimize the cost of the consumed electricity all through the lifetime of the installation. The proper selection of the merit function that must be minimized is of extreme relevance, as different merit functions provide completely different optimum solutions. For instance, a key performance indicator of the system could be its levelized cost of total (consumed) energy or LCOE, which is defined in this work as:

$$LCOE = \frac{CAPEX + \sum_{t'=1}^{T} \left[\dfrac{OPEX(t')}{(1+WACC_{nom})^{t'}}\right]}{\sum_{t'=1}^{T} \left[\dfrac{\sum_{t=1}^{8760}(C_e(t) + C_h(t))\Delta t}{(1+WACC_{real})^{t'}}\right]} \quad (1)$$

where $CAPEX$ is the total capital expenditure (in €), $t'$ is the time variable (in years), $T$ is the total lifetime of the installation (in years), $OPEX(t')$ are the total operational expenditures in year $t'$ (in €), and $\sum_{t=1}^{8760} C_e(t) + C_h(t)\Delta t$ is the total energy consumption (in kWh) during a year, evaluated at time intervals of $\Delta t = 1$ hour. $WACC_{nom}$ and $WACC_{real}$ are the nominal and real weighted average cost of capital, respectively, and are mutually related through $WACC_{real} = (1 + WACC_{nom})/(1 + Infl) - 1$, being $Infl$ the annual inflation rate. The $CAPEX$ of the solution is the addition of the individual $CAPEX$ of each element (PV, HTES, LTES, PGU, Boiler, and THP or EHP), each of which are assumed proportional to the element's nominal capacity. For instance, the $CAPEX$ of the PV system is estimated as $CAPEX_{PV} = CAPEX_{PV}^* P_{nom,PV}$, being $CAPEX_{PV}^*$ the capital expenditure per kW of a PV installation.

A key difference of equation (1) with respect to conventional definitions of LCOE is that in the denominator we put the total energy consumption, rather than the electricity generated by the PV system. Thus, the LCOE defined in equation (1) refers to the cost of the total amount of energy that is consumed (heat plus electricity), including the cost of the electricity and the fuel that are purchased from the grid. The $OPEX(t')$ is calculated as

$$OPEX(t') = (1 + Infl_{el})^{t'} \{OPEX^*_{elec,fix} \max[P_{grid}(t)] +$$
$$OPEX^*_{elec,var} \sum_{t=1}^{8760} P_{grid}(t) \Delta t\} + (1 + Infl_{fuel})^{t'} \{OPEX_{fuel,fix} + \quad (2)$$
$$OPEX^*_{fuel,var} \sum_{t=1}^{8760} Q_{ext}(t) \Delta t\}$$

where $OPEX^*_{elec,fix}$ is the fixed annual cost per installed electric-grid power capacity in current money, and $\max[P_{grid}(t)]$ is the maximum peak-power demanded to the electrical grid, assumed here to be equal to the maximum grid power capacity. This is an important assumption, as the incorporation of storage in the system enables a significant reduction of $\max[P_{grid}(t)]$, subsequently providing a noticeable reduction in the fixed costs of external electric power supply. $OPEX_{fuel,fix}$ is the annual fixed cost of fuel, while $OPEX^*_{elec,var}$ and $OPEX^*_{fuel,var}$ are the annual costs of externally-purchased electricity and heat (in €/kWh$_{el}$ and €/kWh$_{th}$) expressed in current money. Equation (2) implicitly assumes that electricity and fuel prices increase with constant annual energy inflation rates $Infl_{el}$ and $Infl_{fuel}$, respectively (which are not necessarily equal to the overall economy inflation rate $Infl$), and that all years are identical in terms of energy generation and consumption. Other kinds of operational expenditures, such as taxes and the maintenance costs, are neglected.

Preliminary attempts to minimize the total LCOE, as defined in equation (1), resulted in "optimal" solutions tending to maximize the amount of PV electricity dedicated to produce heat. This subsequently resulted in large PV, HTES and LTES systems. Despite the fact that this brings some economical savings after the entire lifetime of the installation, the very high CAPEX results in an intolerable increase of the discounted payback period. As a consequence, we opted for using the levelized cost of consumed electricity (LCOE$_{el}$) as merit function, which is defined in this work as:

$$LCOE_{el} = \frac{CAPEX_t + \sum_{t'=1}^{T}\left[\frac{OPEX_{el}(t')}{(1+WACC_{nom})^{t'}}\right]}{\sum_{t'=1}^{T}\left[\frac{\sum_{t=1}^{8760} C_e(t)\Delta t}{(1+WACC_{real})^{t'}}\right]} \quad (3)$$

and differentiates from equation (1) in that it only considers electricity consumption, i.e.:

$$OPEX_{el}(t') = (1 + Infl_{el})^{t'} \left\{ OPEX^*_{elec,fix} \max[P_{grid}(t)] + OPEX^*_{elec,var} \sum_{t=1}^{8760} P_{grid}(t)\Delta t \right\} \quad (4)$$

The minimization of $LCOE_{el}$ will guide us towards the best system configuration in terms of maximum savings of grid-electricity, which is the most relevant contributor to the total OPEX. Minimizing $LCOE_{el}$ appears to be a more reasonable approach that produces shorter payback periods at the expense of obtaining slightly higher LCOE. The search for the minimum $LCOE_{el}$ is performed in this work by means of the multi-variable direct search (Nelder-Mead) algorithm [57] evaluated over a matrix of different initial conditions to avoid local minimums.

Finally, the discounted payback period is calculated in this study by computing the annual discounted saves (in current money) in the OPEX due to the reduced consumption of external electricity and fuel, with respect to the reference case in *Figure 1*-a. The discounted payback period represents the time needed for these cumulative savings to pay-off the higher CAPEX of the PV+PHPS installation.

All the variables used in this study to describe the PHPS+PV solution are summarized in Table 1. A selected number of them have been used to define the four different economic scenarios that are summarized in Table 2. These scenarios are ordered from more favourable (Scenario 1) to less favourable (Scenario 4). The most favourable scenario assumes a PV CAPEX of 900 €/kW, while the rest of scenarios assume a price of 1200 €/kW (taxes included). These data are selected based on the estimations of the European JCR, according to which the worldwide average price of a residential PV systems without tax was 1150 €/kW[58] in 2018, being the minimum prices found in Australia (950 €/kW). Taking into account that the average learning curve of CAPEX for residential PV installations is in the range of 80-90% [59], it is expectable that CAPEX values below 900 € (taxes included) could be reached in the near future.

Concerning the $CO_2$-equivalent emissions reported in Table 1 it must be noticed that the value for solar PV (20 $gCO_2eq/kWh_{th}$) is taken from the harmonization of a number of published life-cycle assessments conducted by Louwen et al. that assumes a performance ratio of 0.75 and insolation conditions of 1700 $kWh/m^2$-year [60], which are similar to those existing in Madrid. In the case of natural gas, the $CO_2$-equivalent emissions are typically reported in the range of 220 – 280 $gCO_2eq/kWh_{th}$ [61]; thus, we set a value of 250 $gCO_2eq/kWh_{th}$. However, it is worth mentioning that a controversy exists on the determination of the emissions for natural gas in $CO_2$-equivalent units. Natural gas is largely composed of methane, which has a lower atmospheric lifetime (~ 12 years) than $CO_2$ (> 100 years), but a much higher greenhouse potential. Thus, the amount of emissions of natural gas in $CO_2$-equivalent units depends on the time-frame considered for the calculation. Howarth [62] estimated emissions in the range of 550 – 750 $gCO_2eq/kWh_{th}$ for natural gas when considering a shorter timeframe of 20 years. These considerations are not taken into account in this study to keep the coherence among the data obtained from different sources. Finally, the emissions per kWh$_{el}$ of consumed grid electricity is set to 340 gCO$_2$eq/kWh$_{el}$, which is the one obtained for Spain in 2013 from a Well-To-Wheels (WTW) analysis that takes into account not only the direct emissions in the generation site, but also the upstream emissions associated with the fuel extraction and transport, and the power losses along the grid [63].It is worth mentioning that the average emissions for EU grid electricity is 447 gCO$_2$eq/kWh$_{el}$, meaning that a PHPS system installed in Spain has a comparatively lower potential to reduce the CO$_2$ emissions than in other countries such as Germany (615 gCO$_2$eq/kWh$_{el}$) or Italy (431 gCO$_2$eq/kWh$_{el}$). The investigation of the impact of this solution on different emplacements is not the aim of this study.

It is also worth commenting on the assumption of an identical CAPEX for both EHP and THP. Some studies have reported air conditioning prices ranging from 500 to 700 €/kW$_{cool}$ [64]. These values are in agreement with the empiric relation between the capital cost and the heating capacity (kW$_{heat}$) for EHPs reported by Staffell in 2012 as £$_{2012}$/kW$_{heat}$ = 200 + 4750/kW$_{heat}^{1.25}$[7]. This equation results in CAPEX of 395 and 590 €$_{2012}$/kW$_{heat}$ for systems with large (20 kW$_{heat}$) and medium (10 kW$_{heat}$) heating capacities, respectively. Unfortunately, the price of THPs is not as well reported in the literature. In 2002, Grossman [65] estimated the cost of LiBr-water absorption chiller technology in the rage of 165 – 200 $/kW$_{cool}$. However, in 2008, Kim and Infante Ferreira [66] reported significantly higher prices (300 - 400 €/kW$_{cool}$) for the same kind of technology. In both studies, it is estimated that the higher COP of double- and triple-stage chillers would result in lower prices per unit of cooling capacity, despite the higher complexity. Although a lower CAPEX has been reported for THPs, in this study we set an identical price for both EHP and THP of 500 €/kW$_{cool}$ in order to cope with the uncertainty in the published data.

A constant COP is assumed for the EHP all through the year. A value of 4 has been used in most of the cases, as this is the average seasonal energy efficiency ratio (SEER) for residential air conditioners installed worldwide, as reported by the IEA in 2019. However, the best SEER available in the market are up to ~ 12. According to the IEA, it is unlikely that these efficient devices reach the market in many countries, but it illustrates the huge potential for improvement in the near future. For the THP, current absorption technology's COPs range from 0.4 – 1.7 depending on the operation temperature and number of stages. Low temperature (75 – 90 ºC) single-stage chillers have COPs in the range of 0.4 to 0.7; double-stage chillers operating at intermediate temperatures of ~ 150 ºC have COPs in between 1.2 and 1.4; and high temperature (200 – 250 ºC) triple-stage chillers have COPs of up to ~ 1.7. [35], [66], [67]. The COP for THP will be set to 1.3; thus, representing the case of a double-stage chiller operating at intermediate temperatures of ~ 150 ºC. This temperature is compatible with typical indirect pressurized hot water storage systems. Thus, for the case of PHPS-T systems we will consider the combination of COP = 1.3 and a pressurized hot water store at 150 ºC. As it will be seen in the discussion section, lower COPs do not provide a significant advantage with respect to the PHPS solution comprising an EHP.

Concerning the energy prices, two main components have been assumed: energy and network. Typically, energy and taxes are charged per kWh of consumed energy, while the network component is charged annually per contracted kW$_{el}$. The share of the fixed and variable components in the price of natural gas and electricity varies significantly among countries . The increasing amount of grid-connected distributed renewable power systems is triggering a trend to increase the fixed components of the price [68], which contributes to increase the uncertainty of this variable in the future. In this study we assume a price of electrical energy and power of 0.17 €/kWh$_{el}$ and 50 €/kW$_{el}$-year, respectively, which results in an average annual price of electricity of 22.41 c€/kWh$_{el}$ for the specific consumption profile of the reference case considered in this study (*Figure 1*-a). This value is only slightly below the average price of electricity for households' consumers in Spain, as reported by Eurostat, of 24.77 c€/kWh$_{el}$. The resultant share of fixed component in the grid electricity price is 24%. For the natural gas, we assume a price for heating energy and power of 0.07 €/kWh$_{th}$ and 60 €/year, respectively, which results in an average price of 7.39 c€/kWh$_{th}$ for the specific consumption profile of the reference case in *Figure 1*-a. This price is slightly below the average value reported for Spain in Eurostat of 8.75 c€/kWh$_{th}$, and the fixed term represents only 5% of the total price.

## 3. Results and Discussion

This work aims at determining under which circumstances a PHPS system could be profitable for the self-consumption of solar PV electricity in the residential sector. To afford this analysis, we will search for the optimal sizing of the three main elements of the system, i.e. the storage capacity of HTES in kWh$_{th}$, the peak power output of the PGU in kW$_{el}$, and the nominal power of the solar-PV installation in kW$_{el}$, which result in the minimum costs of the consumed electricity after the entire lifetime of the installation. This optimal sizing depends on a number of parameters, such as the cost of grid-electricity and fuel, or the cost and productivity of the PV system, among many others. In this study, we will pay special attention to those parameters related to the PGU and HTES components, as they are not typically used in residential applications and subsequently, there are no reliable data on their cost and performance. Such elements are described in this study by the following five parameters: the PGU conversion efficiency ($\eta_{PGU}$), the HTES CAPEX ($CAPEX^*_{TES}$ in €/kWh$_{th}$), the HTES thermal losses ($a_{HTES}$ and $\Delta T_{HTES}$), and the PGU CAPEX ($CAPEX^*_{H2P}$ in €/kW$_{el}$). Performing a bottom-up estimation of these five parameters would be very unprecise, provided the immaturity of these technologies; thus, our first analysis focuses on conducting an up-bottom analysis, according to which the minimum value of $\eta_{PGU}$, and the maximum values of $CAPEX^*_{HTES}$, $CAPEX^*_{PGU}$, $a_{HTES}$ and $\Delta T_{HTES}$ that leads to profitability will be determined assuming a favourable economic scenario. Thus, the outcome of this analysis will be the technological requirements for each element of the solution in order to reach profitability. This information will be useful to assess the candidate technologies. Nevertheless, despite the fact that we will mention a few possible specific options, it is not the aim of this study to provide a thorough assessment of the most suitable technological implementation.

To quantify the "profitability" of the solution we will look at the amount of electricity savings enabled by the PHPS system. Electricity savings can only be attributed to the presence of significantly large PV, HTES and PGU systems (optimized variables) that provide a significant reduction of the cost of consumed electricity (merit function for the optimization). The first part of the discussion focuses on the PHPS-E configuration (*Figure 1*-b), which is characterized by using an electrically driven heat pump for cooling; thus, the rejected heat from the PGU is only used to satisfy the DHW and space heating needs. In the second part of the discussion, we assess the PHPS-T configuration (*Figure 1*-c), and we discuss under which conditions a PHPS-T system would be preferable than a PHPS-E one. Some of the most relevant results are summarized in Table 3 in order to facilitate the readability of the analysis. Table 4 shows identical results but for the ideal case of loss-less (adiabatic) HTES and LTES systems; thus, they represent the upper bounds of performance for the PHPS solution, which are unattainable in practice.

### 3.1. PHPS-E system

*Figure 3* (b-d) shows the savings on grid electricity resulting from an optimized PHPS-E system as a function of different variables. Every dot in these figures represents an optimized system, meaning that the optimal sizing of the three main elements of the system (PV, HTES and PGU) are set to minimize the cost of the electricity consumed during the system lifetime. The rest of parameters are set to the values reported in Table 1, and to those corresponding to the most favourable scenario in Table 2 (Scenario 1). In the case of *Figure 3*-b, the two independent variables are the PGU efficiency and the HTES and PGU CAPEX. As it could be expected, the largest savings are obtained for high PGU efficiency and small HTES and PGU CAPEX. Besides, the minimum attainable savings of electricity are about 30%, and correspond to the case of direct self-consumption of PV electricity, without storage. The results shown in *Figure 3*-b enable determining the maximum cost targets for HTES and PGU technologies if the PGU conversion efficiency is known. Similarly, it is possible to estimate the minimum PGU conversion efficiency that is needed to reach profitability if the costs of HTES and PGU devices are provided. For instance, a CAPEX of ~ 300 €/kWh$_{th}$ and ~ 4000 €/kW$_{el}$ for the HTES and PGU devices are tolerable (enable a significant savings on electricity consumption) if the PGU conversion efficiency is higher than ~ 40%. However, if the conversion efficiency is 20%, the maximum tolerable CAPEX is reduced to ~ 100 €/kWh$_{th}$ and ~ 2000 €/kW$_{el}$, respectively.

It is worth noticing that high thermal-to-electric conversion efficiencies (> 40%) and low costs (< 1000 €/kW) are attainable by current state of the art Rankine and open-cycle Brayton engines of large (> 1 MW$_{el}$) [39], [69], [70] and medium (30 - 300 kW$_{el}$) [71] sizes. Unfortunately, current state of the art small-scale (< 10 kW$_{el}$) closed-cycle engines, which have been developed for both terrestrial and space-power applications, have either low conversion efficiency (e.g. Rankine/ORC) or high cost (Brayton) [72], [73]. Probably, among all the current dynamic-engine options, the best choice is a Stirling engine, which has a conversion efficiency in the range of 30 - 40% at power outputs ranging from 1 to 30 kW$_e$ [73]. However, low power costs (~ 2,000 €/kW$_{el}$) are only attainable by the largest units (~ 30 kW$_{el}$), whilst the smallest units (~ 1-2 kW$_{el}$) could reach power costs more than 10,000 €/kW$_{el}$. This price lies outside the profitability limit, as seen in *Figure 3*-b. Solid-state converters are better suited for power generation at small-scale and could lead, in principle, to low power costs. In this regard, thermoelectric generators are the most mature technology, but they lack of high conversion efficiency (typically below ~ 10%). A particularly interesting highly-efficient alternative is thermophotovoltaics (TPV), which has been already assessed theoretically for PHPS applications [41], [46]. Despite its much lower degree of development, TPV has already demonstrated significantly higher conversion efficiencies (~ 24%) [74], becoming the most efficient solid-state thermal-to-electric converter to date. TPV is particularly well suited for ultra-high temperature (> 1000 ºC) heat conversion; thus, enabling the use of ultra-dense heat stores, such as silicon or boron latent heats [41]. TPV technology has also potential to reach very low power costs, even below 300 €/kW$_{el}$ [46]. Thus, in the opinion of the authors, TPV should be regarded as a promising choice for future developments in residential (small-scale) PHPS solutions.

*Figure 3*-c illustrates how heat losses in the HTES, i.e. the parameters $a_{HTES}$ and $\Delta T_{HTES}$, impact on the profitability of the PHPHS-E solution. As it could be expected, small $a_{HTES}$ and $\Delta T_{HTES}$ values (small heat losses in the HTES) result in larger savings in electricity. For instance, reducing the HTES temperature below ~ 500 ºC allows a significant reduction of the heat losses, or alternatively enables the use of higher $a_{HTES}$ values; thus, simplifying the thermal insulation design, and eventually enabling a reduction of its cost. However, lower HTES temperatures could result in lower PGU conversion efficiencies and negatively impact on the profitability of the solution. For instance, a HTES operating at 1500 ºC with 40% conversion efficiency requires $a_{HTES} < 0.2$ for reaching electricity savings more than ~ 80%. Reducing the HTES temperature to 250 ºC while keeping a conversion efficiency of 40% could enable similar

savings for less a restrictive $a_{HTES} < 0.4$. However, if the efficiency drops to 20%, such amounts of savings would not be reachable even using an ideal thermal insulation system ($a_{HTES} = 0$). This illustrates an existing trade-off on the optimal HTES temperature between the PGU conversion efficiency and the HTES's thermal insulation losses. High conversion efficiencies and low thermal losses are needed to reach profitability, but both of them increase with temperature. Nevertheless, a very interesting result illustrated in *Figure 3*-c is that this trade-off disappears for temperatures more than ~ 500 ºC. For such high temperatures, $a_{HTES}$ values near or below ~ 0.1, and PGU efficiency more than ~ 20% result in a profitable solution, independently of the HTES temperature. This is partially attributed to the higher storage energy density that is attainable at very high temperatures, which enables a more compact and smaller HTES that compensates the larger amounts of heat losses per unit of store area at higher temperatures. According to this result, the HTES temperature could be as high as required to reach high energy density and PGU conversion efficiency, provided that a low-cost thermal insulation system with $a_{HTES}$ values near or below ~ 0.1 are attainable at such temperatures. Therefore, the selection of the optimal operation temperature should be made based on an overall techno-economic analysis of the entire solution, being the main objective the achievement of high PGU efficiency and low thermal insulation losses at low cost. In this regard, the results shown in *Figure 3*-c could guide the design of an optimal thermal insulation system for PHPS applications.

The results in Table 4 show to the ideal case of adiabatic HTES and LTES systems (i.e. $a_{HTES} = a_{LTES} = 0$). Despite the fact that this ideal situation is unattainable in practice, these results are valuable to set the upper bounds of the technology. If compared with the results in Table 3, (which assume $a_{HTES} = a_{LTES} = 0.1$) the self-consumption ratio is drastically increased due to the reduced amount of heat losses. This results in a smaller PV system and CAPEX, which leads to significantly shorter payback periods. These results highlight the relevance of thermal insulation in the PHPS solution, at least for small-scale residential applications.

*Figure 3*-d assess the sensibility of the PHPS-E solution to variations in the economic conditions. The results shown in *Figure 3*-c and d assume the most favourable case, which is characterized by low PV CAPEX (900 €/kW$_{el}$), low WACC$_{nom}$ (2%), and a high energy price inflation rate (*Infl*$_e$ = 4%). This optimistic scenario allows estimating the minimum requirements for the PHPS-E solution (in terms of cost and heat losses of the HTES, and efficiency of the PGU) to reach profitability in the most favourable economic conditions. *Figure 3*-d illustrates how these requirements are modified for more unfavourable economic scenarios. By calculating the savings on consumed grid electricity resulting from an optimized PHPS-E system as a function of the *Infl*$_e$ and the WACC$_{nom}$ we observed that the scenarios having the same

difference between the $WACC_{nom}$ and the energy price inflation rate result in identical optimal systems and savings of electricity. This means that the profitability can be analyzed through a single variable: the difference between the $WACC_{nom}$ and $Infl_e$. *Figure 3*-d illustrates the combined effect on the electricity savings of the PV CAPEX and ($WACC_{nom}$ - $Infl_e$). As expected, having a low PV CAPEX and an energy inflation rate higher than the $WACC_{nom}$ is very important to reach profitability, especially when the PGU conversion efficiency is low. Higher PGU conversion efficiencies enable reaching profitability under less favourable economic conditions. However, even in the case of high PGU conversion efficiency (e.g. 40%), a low PV CAPEX (below 1000 €/kW$_{el}$) and an energy price inflation rate greater than $WACC_{nom}$ are needed to reach electricity savings of 80% or beyond. A less favourable scenario, e.g. PV CAPEX of 1300 €/kW$_{el}$ and an energy price inflation rate 1% smaller than the $WACC_{nom}$, would enable maximum electricity savings of ~ 55% if the PGU conversion efficiency is 40%, and of ~ 30% if the PGU conversion efficiency is 20%. The latter represents a solution without HTES nor PGU devices, in which the PV electricity is directly consumed without being stored. This illustrates how important are the economic boundary conditions on the potential of the PHPS-E concept.

All the results shown in *Figure 3*(b-d) show deviations of a selected number of parameters (i.e. CAPEX of HTES, PGU and PV, PGU efficiency, $WACC_{nom}$ and energy price inflation rate) from the most favourable economic scenario indicated in Table 2. Besides, they illustrate the electricity savings only, as they represent a quantification of the profitability of a PHPS solution. But no observations have been made so far on other relevant parameters such as the discounted payback period, the savings of fuel and $CO_2$ emissions, or the self-consumption ratio, among others. In the next discussion, we will pay attention to these parameters. To that end, Figure 4 shows a selection of eight system's parameters as a function of the PGU conversion efficiency, for an optimized PHPS-E system under the four different economic scenarios indicated in Table 2. The selected parameters for the representation are: the optimal PGU maximum power capacity (a), the optimal HTES storage capacity (b), the optimal PV nominal power (c), the discounted payback period (d), the savings in fuel (e), and $CO_2$ emissions (f), the self-consumption ratio (g), and the savings in electricity (h). In these graphs, each dot represents the best solution obtained by the direct-search algorithm after being evaluated with 78 different initial simplexes to avoid local minimums.

The first important observation is that there exists a threshold value for the PGU conversion efficiency beyond which the PHPS-E system becomes profitable, i.e. the optimal sizes for the PGU (Figure 4-a) and the HTES (Figure 4-b) are significant. Besides, this threshold efficiency is higher when the economic conditions are less favourable. For instance, the most favourable scenario (Scenario 1) enables profitability at PGU conversion efficiencies more than ~ 10%, while the most unfavourable one (Scenario 4) requires PGU conversion efficiencies more than ~ 80%. For efficiencies below those thresholds, the optimal sizes for the PGU and HTES devices tend to zero, and the only component that remains in the solution is the PV system. In this case, the PV electricity is consumed instantaneously (without storage in the HTES); thus, resulting in a small optimal PV system that enables very high self-consumption ratios (~ 80%, Figure 4-g). Notice that such high self-consumption ratios are attributed to the use of solar PV electricity to produce low-grade heat that is stored in the LTES. This explains why such high self-consumption ratios are combined with small electricity savings (~ 30%, Figure 4-h).

When the PGU efficiency is higher than the profitability threshold, the optimal PHPS-E system comprises significantly large HTES, PGU and PV systems (Figure 4-a, b, c), subsequently enabling electricity savings in the range of 70 - 95% (Figure 4-h), depending on the specific economic boundary conditions. The most favourable scenarios bring particularly large optimal HTES (up to 80 kWh$_{th}$), PGU (up to 3 kW$_{el}$) and solar PV (up to 12 kW$_{el}$) systems, along with higher electricity savings (up to 95%). On the other hand, less favourable economic conditions bring smaller optimal HTES (10 – 20 kWh$_{th}$), PGU (~ 0.8 kW$_{el}$), and solar PV (5 – 7 kW$_{el}$) systems, as well as lower electricity savings of 70 – 75%. A particularly important observation is that increasing the efficiency beyond a certain value does not improve the savings in electricity. Instead, it leads to lower savings of fuel and subsequently higher $CO_2$ emissions. This critical efficiency is ~ 40% for the scenarios 1 and 2, and ~ 60% (~ 80%) for the scenario 3 (4). At these specific efficiencies, the $CO_2$ emissions are minimum, as higher PGU conversion efficiencies result in the production of smaller amounts of heat that bring a larger consumption of fuel. Unfortunately, the same efficiency that minimizes the $CO_2$ emissions maximizes the payback period and minimizes the self-consumption ratio, as it implies the use of significantly large PV, HTES and PGU systems. However, the potential reduction in the payback period due to an increment of the conversion efficiency beyond such values might not be very significant. For instance, in the Scenario 1, the payback period is reduced only 3 years (from 14 to 11 years) when the PGU conversion efficiency increases from 40% to 80%, while the electricity savings are not improved. This illustrates that increasing the PGU conversion efficiency beyond a certain critical value (which depends on the boundary economic conditions) might not be as important as it could be intuitively expected.

Other interesting observation concerns the very small self-consumption ratios (35 – 60%) that are obtained, meaning that a high amount of PV electricity is finally wasted. To understand the source of these losses, *Figure 5* shows the four main contributions to the energy lost in the system as a function of the PGU conversion efficiency for each of the four economic scenarios indicated in Table 2. For low conversion efficiencies, i.e. when the solution lacks of both HTES and PGU devices, most of the losses (> 80%) are attributed to heat losses in the LTES.

However, these losses are not very significant in absolute value, as they enable very high self-consumption ratios of ~ 80% (Figure 4-g). On the other hand, when the PGU conversion efficiency is large enough, and subsequently the optimal HTES and PGU sizes are significant, the HTES thermal insulation heat losses account for most of the energy losses of the PHPS-E solution (70 – 90%). Thus, heat losses in the HTES are the main reason for the low self-consumption levels observed in Figure 4-h. The next contribution is the heat lost in the LTES (10 – 20%). Both the PV electricity that is neither consumed nor stored, and the exhaust heat from the PGU converter that is wasted, represent a negligible contribution to the total energy losses in the system. Only in Scenario 1 the PV electricity that is directly wasted could represent a significant amount of losses (up to 20%) due to the oversized PV installation, which is possible due to its low cost. But even in this case the heat losses in the HTES represent the highest contributor to the overall energy losses of the system. Therefore, improving thermal insulation of the HTES is key to improve the self-consumption ratio. However, it is important to notice that, even with such low self-consumption ratios, the PHPS-E system can provide significant electricity savings (> 70%) and reach profitability with reasonably short payback periods. The heat losses through the HTES thermal insulation system could be recovered as low-grade heat to further reduce the amount of fuel consumption, as proposed in [47], [48]. Analysing this and other possible improvements could be the aim of a future work.

### 3.2. PHPS-T system

This last part of the article focuses on the analysis of the PHPS-T system (*Figure 1*-c), where a thermally-driven heat pump (THP) is used instead of an electrically-driven one to satisfy the cooling demand. In this case, the heat generated in the PGU is not only used for DHW and space heating, but it is also used for powering the THP and satisfying the cooling needs in summer season. Thus, one could expect a more efficient use of the generated heat all through the year. However, one could also argue that an increment in the COP of the EHP could have a similar effect on reducing the electricity consumption both in the reference case (*Figure 1*-a) and in the PHPS-E solution (*Figure 1*-b); thus, hindering the profitability of the PHPS-T solution. Thus, the payback period of the PHPS-T system should be analysed as a function of the difference between the COPs of the THP and the EHP.

In this regard, Figure 6 shows the payback period of an optimized PHPS-T (top), and the difference between the payback periods of PHPS-E and PHPS-T solutions (bottom) as a function of the THP and EHP COPs for four different PGU conversion efficiencies. Payback periods are calculated with respect to the reference case (Figure 1-a). As expected, increasing the EHP COP produces a significant increment on the payback period of the PHPS-T solution, hindering its profitability (Figure 6-top). This increment could result in intolerable payback periods (> 20 years) if the THP COP is low (< 0.7) and the EHP COP is reasonably high (> 4). In this case, the PHPS-E solution would be the preferable solution, with payback periods significantly shorter than those of PHPS-T. Very high THP COPs (more than ~ 1.3) would be needed for PHPS-T to reach reasonably low payback periods (< 12 years) when the EHP COP is considerably high (~ 6 or beyond). In this case, the PHPS-T solution could provide significantly shorter payback periods than PHPS-E. It is worth noticing that the improvement of the EHP COP is beneficial not only for the PHPS-E system, but also for the reference case. Thus, improving the EHP COP does not bring a significant reduction of the payback period for the PHPS-E system with respect to the reference case. For this reason, the only way to drastically reduce the payback period is to adopt the PHPS-T solution with a very highly efficient THP. High THP COPs of 1.2 – 1.7 can be attained with current state of the art double-stage or triple-stage absorption chillers operating at temperatures in the range of 150 - 250 ºC [35], [66], [67]. This implies that a PHPS-T solution should rely on the use of PGU with a high-grade rejected heat (> 150 ºC) able to power an efficient double- or triple-stage absorption chiller. Otherwise, the PHPS-E system would be a preferable solution. This will impact on the selection of the PGU. High rejection temperature dynamic engines or solid-state devices are preferable. In this regard, the low rejection temperature of TPV devices is detrimental. High TPV cell temperatures lead to a significant reduction of the conversion efficiency. Thus, the requirement of higher temperature rejected heat points in the direction of an interesting research line for TPV, which is the development of highly efficient TPV cells able to operate at temperatures of ~ 200ºC [72]. Other high-rejection temperature solid-state alternatives, such as thermionic generators [72], could also be regarded as an interesting option for the future.

Figure 7 shows the optimization results for a PHPS-T system with $COP_{EHP} = 4$ and $COP_{THP} = 1.3$; thus, representing a case where the PHPS-T system provides shorter payback periods than PHPS-E, as shown in Figure 6 (bottom). The results are presented in a similar way than it was done in *Figure 4* for the PHPS-E system, evaluating the four scenarios in Table 2. Despite the fact that the general tendencies are similar, there are some remarkable differences. First, the optimal PHPS-T system requires smaller components, ultimately resulting in significantly shorter payback periods (8 – 14 years). This is mostly attributed to the reduction in the electricity consumption when removing the EHP, which results in a significantly lower amount of electricity that needs to be generated and stored by the PV+PHPS system; thus, leading to smaller HTES and PGU devices. It is worth mentioning that the smaller HTES and PGU devices also result in a slightly lower potential of the PHPS-T solution to reduce the grid-electricity consumption. This illustrates an existing trade-off between obtaining either small payback periods or high energetic savings.

The other main difference observed when comparing PHPS-T (*Figure 7*) with PHPS-E (*Figure 4*) is the increment in the consumption of fuel (up to ~ 20%). The THP consumes large amounts of heat, especially in summer, which cannot be fully provided by the rejected heat from the PGU. This implies that a larger amount of external fuel is consumed. This contributes to the increase of the $CO_2$ equivalent emissions, which are barely compensated by the savings in emissions due to the self-consumption of solar-PV electricity. As a result, the PHPS-T solution produce significantly lower savings in $CO_2$ emissions (up to ~ 0.75-ton eq$CO_2$/year, from a total emission of 5.2 ton/year). If the $COP_{THP}$ would be lower, a PHPS-T solution could even produce an increment of the emissions with respect to the reference case. It must be noticed that these results are obtained for a system that aims at minimizing the lifetime cost of electricity. Choosing other kinds of merit functions (e.g. the total LCOE in eq. 1) would lead to larger HTES and PGU systems that bring higher savings in $CO_2$ emissions at the expenses of increasing the payback period. Another obvious strategy to reduce the $CO_2$ emissions could be hybridizing a PHPS-T solution with a kind of renewable heat source, such as solar thermal collectors. The resultant hybrid PHPS-T solution would enable a significant reduction in the $CO_2$ emissions of up to 4.5 ton/year (from the total reference emissions of 5.2 ton/year), as well as a low payback period of 12 years or below.

## 4 Conclusions

In this article we have evaluated the implementation of a power-to-heat-to-power storage solution for the self-consumption of solar PV electricity in a dwelling in Madrid. Our results indicate that the solution has potential to provide a significant amount of savings in the consumption of grid electricity (> 70%) with reasonably short discounted payback periods (< 15 years), if compared to a solution that uses fuel for heating and grid-electricity. This holds true even when the heat-to-power conversion efficiencies is moderately low (20 - 30%), provided that the economic conditions are favourable. If the cost of the technology is sufficiently low, there exists a certain thermal-to-electric conversion efficiency (~ 40% in the most favourable case analysed in this study) beyond which the power generation system can be made slightly smaller, but the amount of savings in electricity does not increase, i.e. increasing the efficiency might not be as important as it could be intuitively expected. Besides, the inefficient power conversion process results in a large amount of rejected heat that can be used to satisfy other kinds of energetic demands. For instance, if an electrically driven heat pump is used for cooling, the rejected heat could be used for providing space heating and domestic hot water, bringing additional savings in the fuel consumption in the range of 10 – 20%, and global savings in $CO_2$ equivalent emissions in the range of 1.2 – 1.6 ton/year (24 – 31%). If a thermally driven heat pump is used instead, the generated heat could be also used for satisfying the cooling demands. This leads to shorter payback periods if the COP of the thermally driven heat pump is significantly high (> 1.3). This condition is attainable by current state of the art double- or triple-stage absorption chillers operating at temperatures in the range of 150 – 250 ºC, meaning that a high rejection temperature would be necessary in the conversion of the stored heat into electricity. However, the very large cooling needs during summer in Madrid may not be fully satisfied by the rejected heat in the system, in which case an extra consumption of fuel is needed, subsequently leading to a small reduction (or even an increment) in the greenhouse gas emissions. A possible solution consists of replacing the use of fossil fuels by a renewable heat source, such as solar thermal. In this case, the entire power-to-heat-to-power solution comprising a highly efficient thermally driven heat pump and solar thermal collectors could provide a drastic reduction of the emissions (~ 4.5 ton/year, or 86%) while keeping reasonably short payback periods (< 12 years). Notice that a clearer economic advantage would be obtained if we compared the proposed PHPS solution with a reference case that uses an electric-powered boiler, instead of a fuel-powered one. This most favourable scenario is not analysed in this study, and should be assessed in future works.

The main drawback of the proposed solution, independently of using a thermally- or an electrically-driven heat pump, concerns the small self-consumption ratios of PV electricity (40 – 60%), which are mainly attributed to the large amount of heat losses in the high temperature thermal store. Possible ways of minimizing these losses include the development of novel ultra-dense heat stores at moderately low temperatures, or more advanced thermal insulation systems. Another disadvantage concerns the low readiness level of heat-to-power generation technologies that are highly efficient at small scales (~ 1 kW). Some solid-state power generators, such as thermophotovoltaics, show potential to reach the required conversion efficiency and low cost. However, they are not readily available in the market. Next works should assess the use of PHPS systems in larger residential buildings, where the use of market-available dynamic engines, such as Stirling generators, could be profitable. Future analyses should also assess the possibility of selling the excess of PV electricity to the grid, which has been disregarded in the current work. This could negatively impact on the profitability of the solution as it is considered in this study.

Finally, regardless of the particular system implementation, the system model and methodology presented in this article becomes a quite powerful tool which will allow to answer fundamental questions regarding the profitability of different energy solutions (including novel energy storage technologies) by evaluating payback periods and energy saving as well as optimal sizing that results in minimum costs of energy, among many other parameters.

## 5 Acknowledgements

This work has been partially funded by the projects MADRID-PV2-CM (P2018/EMT-4308), funded by the government of the Comunidad de Madrid with the support from FEDER Funds, TORMES (ENE2015-72843-EXP), and AMADEUS (737054), funded by the European Commission through the Horizon 2020 programme, FET-OPEN action. The sole responsibility for the content of this publication lies with the authors. It does not necessarily reflect the opinion of the European Union. Neither the REA nor the European Commission are responsible for any use that may be made of the information contained therein. A. Datas acknowledges postdoctoral fellowship support from the Spanish "Juan de la Cierva-Incorporación" program (IJCI-2015-23747). A. Ramos acknowledges the Universitat Politècnica de Catalunya for her Serra Hunter Tenure Track professor post. The authors acknowledge the fruitful discussions to Dra. Marta Victoria Pérez, especially concerning the methodology to assess the economic aspects of the system, and to Dr. Stephan Lang for his advices concerning the thermal losses analysis in the heat stores.

**References**

[1] P. Nejat, F. Jomehzadeh, M. M. Taheri, M. Gohari, and M. Z. Abd. Majid, "A global review of energy consumption, CO2 emissions and policy in the residential sector (with an overview of the top ten CO2 emitting countries)," *Renewable and Sustainable Energy Reviews*, vol. 43, pp. 843–862, Mar. 2015.

[2] D. W. Wu and R. Z. Wang, "Combined cooling, heating and power: A review," *Progress in Energy and Combustion Science*, vol. 32, no. 5, pp. 459–495, Sep. 2006.

[3] H. I. Onovwiona and V. I. Ugursal, "Residential cogeneration systems: review of the current technology," *Renewable and Sustainable Energy Reviews*, vol. 10, no. 5, pp. 389–431, Oct. 2006.

[4] H. Cho, A. D. Smith, and P. Mago, "Combined cooling, heating and power: A review of performance improvement and optimization," *Applied Energy*, vol. 136, pp. 168–185, Dec. 2014.

[5] D. Sonar, S. L. Soni, and D. Sharma, "Micro-trigeneration for energy sustainability: Technologies, tools and trends," *Applied Thermal Engineering*, vol. 71, no. 2, pp. 790–796, Oct. 2014.

[6] L. Trygg and S. Amiri, "European perspective on absorption cooling in a combined heat and power system – A case study of energy utility and industries in Sweden," *Applied Energy*, vol. 84, no. 12, pp. 1319–1337, Dec. 2007.

[7] I. Staffell, D. Brett, N. Brandon, and A. Hawkes, "A review of domestic heat pumps," *Energy Environ. Sci.*, vol. 5, no. 11, pp. 9291–9306, Oct. 2012.

[8] E. Bellos and C. Tzivanidis, "Multi-objective optimization of a solar driven trigeneration system," *Energy*, vol. 149, pp. 47–62, Apr. 2018.

[9] A. Ramos, M. A. Chatzopoulou, J. Freeman, and C. N. Markides, "Optimisation of a high-efficiency solar-driven organic Rankine cycle for applications in the built environment," *Applied Energy*, vol. 228, pp. 755–765, Oct. 2018.

[10] A. Kasaeian, G. Nouri, P. Ranjbaran, and D. Wen, "Solar collectors and photovoltaics as combined heat and power systems: A critical review," *Energy Conversion and Management*, vol. 156, pp. 688–705, Jan. 2018.

[11] K. E. N'Tsoukpoe *et al.*, "Integrated design and construction of a micro-central tower power plant," *Energy for Sustainable Development*, vol. 31, pp. 1–13, Apr. 2016.

[12] Y. M. Seshie, K. E. N'Tsoukpoe, P. Neveu, Y. Coulibaly, and Y. K. Azoumah, "Small scale concentrating solar plants for rural electrification," *Renewable and Sustainable Energy Reviews*, vol. 90, pp. 195–209, Jul. 2018.

[13] R. Luthander, J. Widén, D. Nilsson, and J. Palm, "Photovoltaic self-consumption in buildings: A review," *Applied Energy*, vol. 142, pp. 80–94, Mar. 2015.

[14] "REN21. 2018. Renewables 2018 Global Status Report," ISBN 978-3-9818107-0-7, 2018.

[15] "IEA SNAPSHOT OF GLOBAL PHOTOVOLTAIC MARKETS," IEA - International Energy Agency, IEA PVPS T1-33:2018, 2018.

[16] REN21, "RENEWABLES 2018 GLOBAL STATUS REPORT."

[17] A. H. Nosrat, L. G. Swan, and J. M. Pearce, "Improved performance of hybrid photovoltaic-trigeneration systems over photovoltaic-cogen systems including effects of battery storage," *Energy*, vol. 49, pp. 366–374, Jan. 2013.

[18] A. S. Mundada, K. K. Shah, and J. M. Pearce, "Levelized cost of electricity for solar photovoltaic, battery and cogen hybrid systems," *Renewable and Sustainable Energy Reviews*, vol. 57, pp. 692–703, May 2016.

[19] S. Acha, A. Mariaud, N. Shah, and C. N. Markides, "Optimal design and operation of distributed low-carbon energy technologies in commercial buildings," *Energy*, vol. 142, pp. 578–591, Jan. 2018.

[20] Y.-Y. Jing, H. Bai, J.-J. Wang, and L. Liu, "Life cycle assessment of a solar combined cooling heating and power system in different operation strategies," *Applied Energy*, vol. 92, pp. 843–853, Apr. 2012.

[21] A. H. A. Al-Waeli, K. Sopian, H. A. Kazem, and M. T. Chaichan, "Photovoltaic/Thermal (PV/T) systems: Status and future prospects," *Renewable and Sustainable Energy Reviews*, vol. 77, pp. 109–130, Sep. 2017.

[22] A. Ramos, M. A. Chatzopoulou, I. Guarracino, J. Freeman, and C. N. Markides, "Hybrid photovoltaic-thermal solar systems for combined heating, cooling and power provision in the urban environment," *Energy Conversion and Management*, vol. 150, pp. 838–850, Oct. 2017.

[23] M. Herrando, A. Ramos, I. Zabalza, and christos N. Markides, "Energy Performance of a Solar Trigeneration System Based on a Novel Hybrid PVT Panel for Residential Applications," in *Proc. of the Solar World Conference*, 2017.

[24] M. Herrando and C. N. Markides, "Hybrid PV and solar-thermal systems for domestic heat and power provision in the UK: Techno-economic considerations," *Applied Energy*, vol. 161, pp. 512–532, Jan. 2016.

[25] A. Selviaridis, B. R. Burg, A. S. Wallerand, F. Maréchal, and B. Michel, "Thermo-economic analysis of a trigeneration HCPVT power plant," *AIP Conference Proceedings*, vol. 1679, no. 1, p. 100004, Sep. 2015.

[26] C. J. C. Williams, J. O. Binder, and T. Kelm, "Demand side management through heat pumps, thermal storage and battery storage to increase local self-consumption and grid compatibility of PV systems," in *2012 3rd IEEE PES Innovative Smart Grid Technologies Europe (ISGT Europe)*, 2012, pp. 1–6.

[27] J. Tong, Z. Quan, Y. Zhao, G. Wang, J. Cai, and Y. Chi, "The Study on the Performance of Solar Photovoltaic and Air Source Heat Pump Composite Building Energy Supply System," *Procedia Engineering*, vol. 205, pp. 4082–4089, Jan. 2017.

[28] C. Roselli, M. Sasso, and F. Tariello, "Integration between electric heat pump and PV system to increase self-consumption of an office application," *Renew. Energy Environ. Sustain.*, vol. 2, p. 28, 2017.

[29] A. Bloess, W.-P. Schill, and A. Zerrahn, "Power-to-heat for renewable energy integration: A review of technologies, modeling approaches, and flexibility potentials," *Applied Energy*, vol. 212, pp. 1611–1626, Feb. 2018.

[30] J. Li and M. A. Danzer, "Optimal charge control strategies for stationary photovoltaic battery systems," *Journal of Power Sources*, vol. 258, pp. 365–373, Jul. 2014.

[31] R. Thygesen and B. Karlsson, "Simulation and analysis of a solar assisted heat pump system with two different storage types for high levels of PV electricity self-consumption," *Solar Energy*, vol. 103, pp. 19–27, May 2014.

[32] D. Parra, G. S. Walker, and M. Gillott, "Are batteries the optimum PV-coupled energy storage for dwellings? Techno-economic comparison with hot water tanks in the UK," *Energy and Buildings*, vol. 116, pp. 614–621, Mar. 2016.

[33] A. Mariaud, S. Acha, N. Ekins-Daukes, N. Shah, and C. N. Markides, "Integrated optimisation of photovoltaic and battery storage systems for UK commercial buildings," *Applied Energy*, vol. 199, pp. 466–478, Aug. 2017.

[34] O. Schmidt, A. Hawkes, A. Gambhir, and I. Staffell, "The future cost of electrical energy storage based on experience rates," *Nature Energy*, vol. 2, no. 8, p. 17110, Aug. 2017.

[35] R. Z. Wang, X. Yu, T. S. Ge, and T. X. Li, "The present and future of residential refrigeration, power generation and energy storage," *Applied Thermal Engineering*, vol. 53, no. 2, pp. 256–270, May 2013.

[36] M. Zheng, C. J. Meinrenken, and K. S. Lackner, "Smart households: Dispatch strategies and economic analysis of distributed energy storage for residential peak shaving," *Applied Energy*, vol. 147, pp. 246–257, Jun. 2015.

[37] Y. Li, X. Wang, D. Li, and Y. Ding, "A trigeneration system based on compressed air and thermal energy storage," *Applied Energy*, vol. 99, pp. 316–323, Nov. 2012.

[38] D. Teichmann, K. Stark, K. Müller, G. Zöttl, P. Wasserscheid, and W. Arlt, "Energy storage in residential and commercial buildings via Liquid Organic Hydrogen Carriers (LOHC)," *Energy Environ. Sci.*, vol. 5, no. 10, pp. 9044–9054, Sep. 2012.

[39] L. A. Weinstein, J. Loomis, B. Bhatia, D. M. Bierman, E. N. Wang, and G. Chen, "Concentrating Solar Power," *Chem. Rev.*, vol. 115, no. 23, pp. 12797–12838, Dec. 2015.

[40] M. A. Sabiha, R. Saidur, S. Mekhilef, and O. Mahian, "Progress and latest developments of evacuated tube solar collectors," *Renewable and Sustainable Energy Reviews*, vol. 51, pp. 1038–1054, Nov. 2015.

[41] A. Datas, A. Ramos, A. Martí, C. del Cañizo, and A. Luque, "Ultra high temperature latent heat energy storage and thermophotovoltaic energy conversion," *Energy*, vol. 107, pp. 542–549, Jul. 2016.

[42] A. Datas *et al.*, "AMADEUS: Next generation materials and solid state devices for ultra high temperature energy storage and conversion," in *AIP Conference Proceedings*, 2018, vol. 2033, p. 170004.

[43] T. Desrues, J. Ruer, P. Marty, and J. F. Fourmigué, "A thermal energy storage process for large scale electric applications," *Applied Thermal Engineering*, vol. 30, no. 5, pp. 425–432, Apr. 2010.

[44] A. Thess, "Thermodynamic Efficiency of Pumped Heat Electricity Storage," *Phys. Rev. Lett.*, vol. 111, no. 11, p. 110602, Sep. 2013.

[45] R. B. Laughlin, "Pumped thermal grid storage with heat exchange," *Journal of Renewable and Sustainable Energy*, vol. 9, no. 4, p. 044103, Jul. 2017.

[46] C. Amy, H. R. Seyf, M. A. Steiner, D. J. Friedman, and A. Henry, "Thermal energy grid storage using multi-junction photovoltaics," *Energy Environ. Sci.*, Nov. 2018.

[47] A. Robinson, "Ultra-high temperature thermal energy storage. part 1: concepts," *Journal of Energy Storage*, vol. 13, pp. 277–286, Oct. 2017.

[48] A. Robinson, "Ultra-high temperature thermal energy storage. Part 2: Engineering and operation," *Journal of Energy Storage*, vol. 18, pp. 333–339, Aug. 2018.

[49] R. Sioshansi, P. Denholm, T. Jenkin, and J. Weiss, "Estimating the value of electricity storage in PJM: Arbitrage and some welfare effects," *Energy Economics*, vol. 31, no. 2, pp. 269–277, Mar. 2009.

[50] P. Srikhirin, S. Aphornratana, and S. Chungpaibulpatana, "A review of absorption refrigeration technologies," *Renewable and Sustainable Energy Reviews*, vol. 5, no. 4, pp. 343–372, Dec. 2001.

[51] J. P. Kotzé, T. W. von Backström, and P. J. Erens, "High Temperature Thermal Energy Storage Utilizing Metallic Phase Change Materials and Metallic Heat Transfer Fluids," *Journal of Solar Energy Engineering*, vol. 135, no. 3, pp. 035001–035001, May 2013.

[52] *EnergyPlus$^{TM}$*. .

[53] Spanish Government, "Spanish Building Code (Código técnico de la edificación): Orden FOM/1635/2013, de 10 de septiembre, por la que se actualiza el Documento Básico DB-HE «Ahorro de Energía», del Código Técnico de la Edificación, aprobado por Real Decreto 314/2006, de 17 de marzo, Boletín Oficial del Estado.," 2013.

[54] B. Atanasiu, J. Maio, D. Staniaszek, I. Kouloumpi, and T. Kenkmann, "ENTRANZE Project, "Overview of the EU-27 Building Policies and Programs. Factsheets on the nine Entranze Target Countries," 2014.

[55] JRC IPTS, 2012, "Green Public Procurement: Windows Technical Background Report. Windows, Glazed Doors and Skylights," 2012.

[56] "PVsyst Software." [Online]. Available: http://www.pvsyst.com/en/software. [Accessed: 11-Sep-2018].

[57] J. Lagarias, J. Reeds, M. Wright, and P. Wright, "Convergence properties of the Nelder-Mead simplex method in low dimensions," *Siam Journal on Optimization*, vol. 9, no. 1, pp. 112–147, Dec. 1998.

[58] A. JÄGER-WALDAU, "PV Status Report 2018," EUR 29463 EN, 2018.

[59] A. M. Elshurafa, S. R. Albardi, S. Bigerna, and C. A. Bollino, "Estimating the learning curve of solar PV balance–of–system for over 20 countries: Implications and policy recommendations," *Journal of Cleaner Production*, vol. 196, pp. 122–134, Sep. 2018.

[60] A. Louwen, W. G. J. H. M. van Sark, A. P. C. Faaij, and R. E. I. Schropp, "Re-assessment of net energy production and greenhouse gas emissions avoidance after 40 years of photovoltaics development," *Nature Communications*, vol. 7, p. 13728, Dec. 2016.

[61] R. Dones, T. Heck, and S. Hirschberg, "Greenhouse Gas Emissions From Energy Systems: Comparison And Overview," 2004.

[62] R. W. Howarth, "A bridge to nowhere: methane emissions and the greenhouse gas footprint of natural gas," *Energy Science & Engineering*, vol. 2, no. 2, pp. 47–60, 2014.

[63] A. Moro and L. Lonza, "Electricity carbon intensity in European Member States: Impacts on GHG emissions of electric vehicles," *Transportation Research Part D: Transport and Environment*, vol. 64, pp. 5–14, Oct. 2018.

[64] N. Pardo Garcia, K. Vatopoulos, A. Krook Riekkola, A. Perez Lopez, and L. Olsen, "Best available technologies for the heat and cooling market in the European Union," Joint Reserach Center, EUR 25407 EN, 2012.

[65] G. Grossman, "Solar-powered systems for cooling, dehumidification and air-conditioning," *Solar Energy*, vol. 72, no. 1, pp. 53–62, Jan. 2002.

[66] D. S. Kim and C. A. Infante Ferreira, "Solar refrigeration options – a state-of-the-art review," *International Journal of Refrigeration*, vol. 31, no. 1, pp. 3–15, Jan. 2008.

[67] Y. Fan, L. Luo, and B. Souyri, "Review of solar sorption refrigeration technologies: Development and applications," *Renewable and Sustainable Energy Reviews*, vol. 11, no. 8, pp. 1758–1775, Oct. 2007.

[68] E. Vartiainen, G. Masson, and C. Breyer, "The True Competitiveness of Solar PV. A European Case Study," European PV Technology and Innovation Platform (ETIP PV), 2017.

[69] C. K. Ho, M. Carlson, P. Garg, and P. Kumar, "Cost and Performance Tradeoffs of Alternative Solar-Driven S-CO2 Brayton Cycle Configurations," presented at the ASME 2015 9th International Conference on Energy Sustainability collocated with the ASME 2015 Power Conference, the ASME 2015 13th International Conference on Fuel Cell Science, Engineering and Technology, and the ASME 2015 Nuclear Forum, 2015, p. V001T05A016-V001T05A016.

[70] S. Quoilin, M. V. D. Broek, S. Declaye, P. Dewallef, and V. Lemort, "Techno-economic survey of Organic Rankine Cycle (ORC) systems," *Renewable and Sustainable Energy Reviews*, vol. 22, pp. 168–186, Jun. 2013.

[71] P. A. Pilavachi, "Mini- and micro-gas turbines for combined heat and power," *Applied Thermal Engineering*, vol. 22, no. 18, pp. 2003–2014, Dec. 2002.

[72] A. Datas and A. Martí, "Thermophotovoltaic energy in space applications: Review and future potential," *Solar Energy Materials and Solar Cells*, vol. 161, pp. 285–296, Mar. 2017.

[73] A. Mazzetti, M. Gianotti Pret, G. Pinarello, L. Celotti, M. Piskacev, and A. Cowley, "Heat to electricity conversion systems for moon exploration scenarios: A review of space and ground technologies," *Acta Astronautica*, vol. 156, pp. 162–186, Mar. 2019.

[74] D. N. Woolf *et al.*, "High-efficiency thermophotovoltaic energy conversion enabled by a metamaterial selective emitter," *Optica, OPTICA*, vol. 5, no. 2, pp. 213–218, Feb. 2018.

[75] EASE (European Association for Storage of Energy), "Thermal Hot Water Storage." .

[76] L. F. Cabeza, E. Galindo, C. Prieto, C. Barreneche, and A. Inés Fernández, "Key performance indicators in thermal energy storage: Survey and assessment," *Renewable Energy*, vol. 83, pp. 820–827, Nov. 2015.

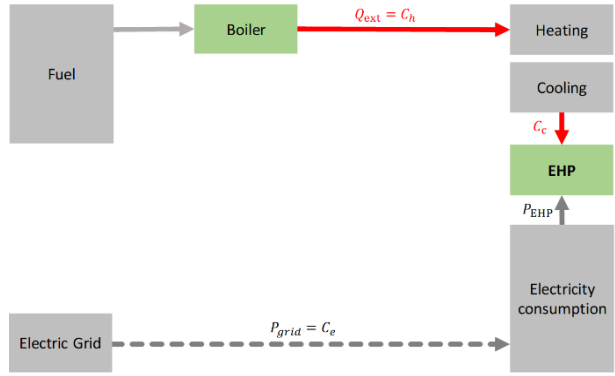

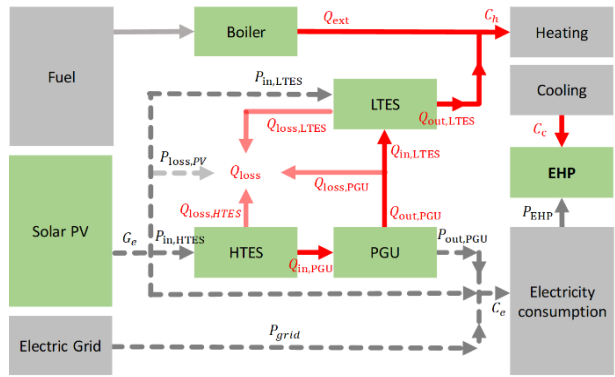

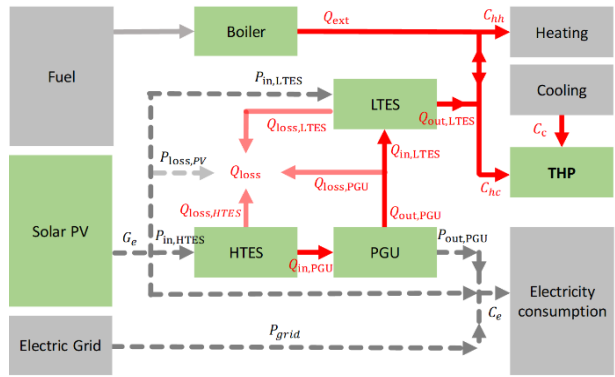

*Figure 1. Illustration of the electrical (grey dashed lines) and thermal (red solid lines) energy flows in the system configurations analysed in this study: (a) reference case comprising a boiler for heating and an electrically driven heat pump (EHP) for cooling, (b) PHPS-E configuration comprising an EHP for cooling, a solar PV system, a high- and low- grade thermal stores (HTES and LTES, respectively), and a power generation unit (PGU), c) PHPS-T configuration similar to the PHPS-E but using a thermally-driven heat pump (THP) instead of an EHP. Notice that in each case, the variables take different values (e.g. the electricity consumption $C_e$ is lower in the PHPS-T system than in the other two cases). The same notation is used to avoid an excessively complicated notation.*

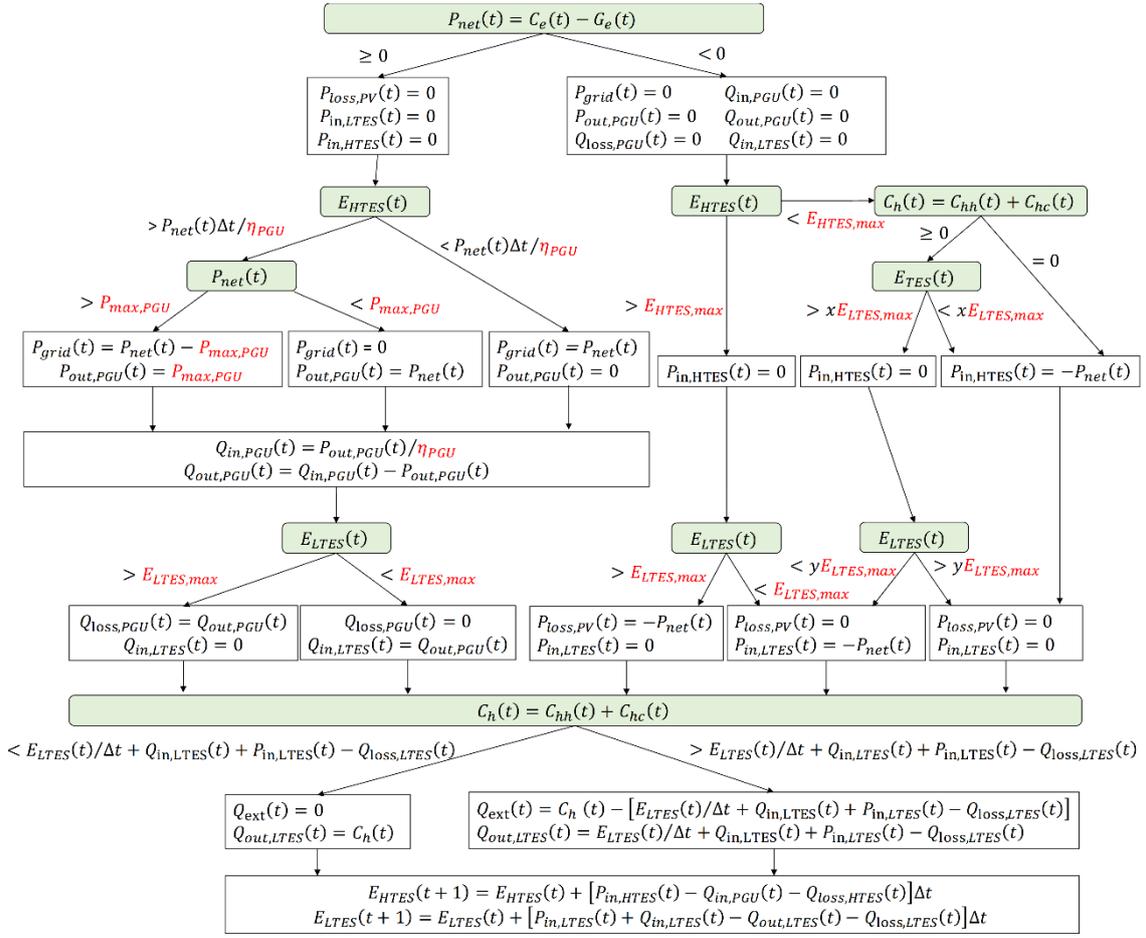

*Figure 2. Energy management algorithm and system model equations.*

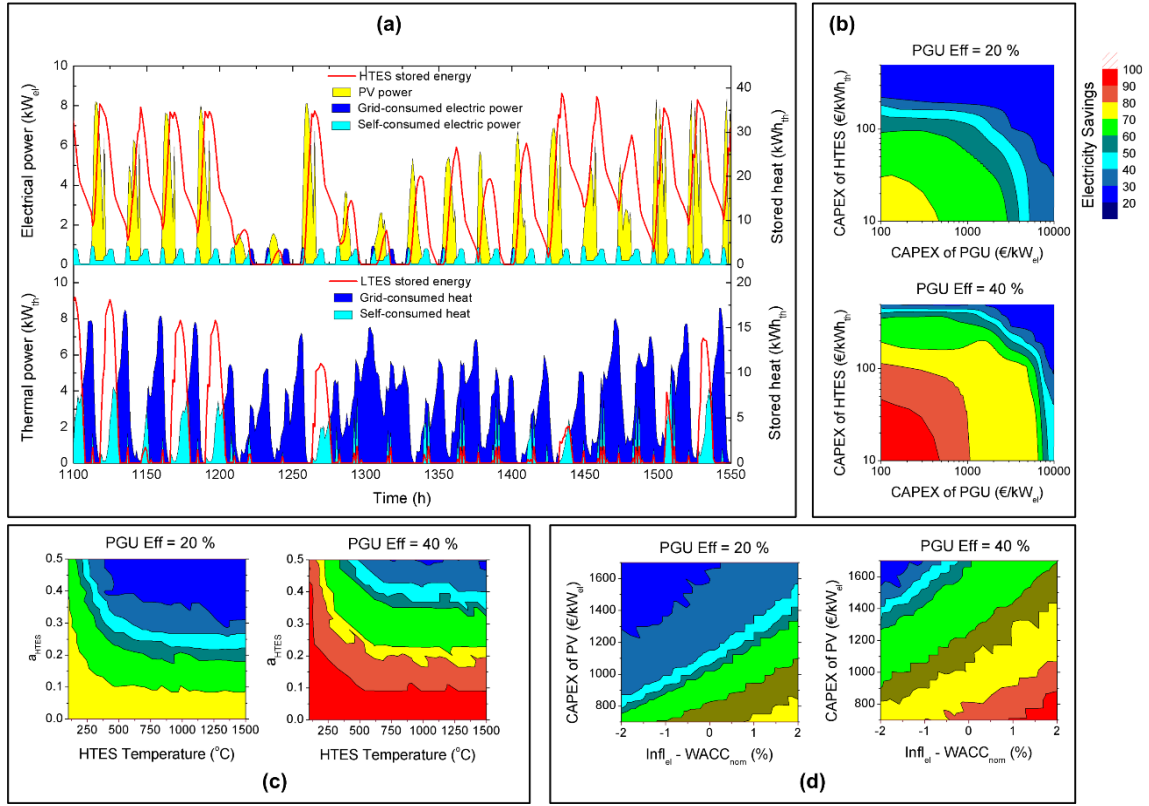

*Figure 3. Simulation results for a PHPHS-E system (Figure 1) under Scenario 1 (Table 1). Other system parameters as set to the values in Table 2. (a) Hourly variation of both grid- and self-consumed electricity along with the thermal energy stored in the HTES (top) and the hourly variation of both grid- and self-consumed heat along with the thermal energy stored in the LTES (bottom) for an optimized system with $\eta_{PGU} = 30\%$, $P_{nom,PV} = 10.6$ kW$_{el}$, $P_{max,PGU} = 1.35$ kW$_{el}$, and $E_{max,HTES} = 33.7$ kW$_{th}$; (b-d) Contour plots representing the savings of electricity, in percentage with respect to the reference case (Figure 1), for an optimized system and two PGU conversion efficiencies (20 % and 40 %) as a function of (b) the CAPEX of PGU and HTES, (c) the HTES temperature and $a_{HTES}$, (d) the CAPEX of PV system and the difference between WACC and the energy price inflation rate.*

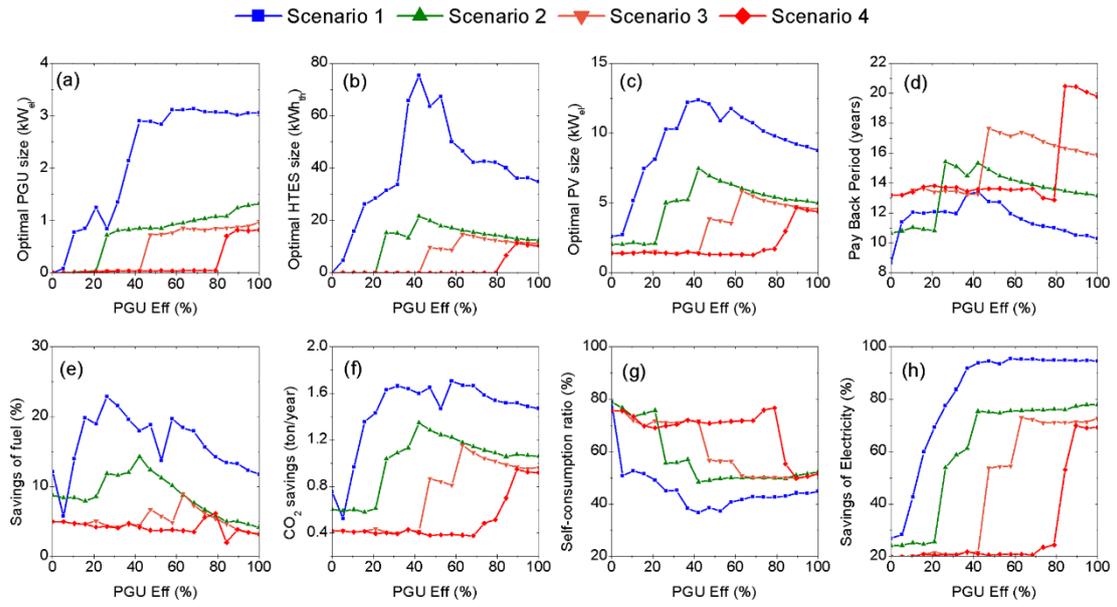

*Figure 4. Performance parameters of an optimized PHPS-E system (Figure 1-b) installed in Madrid as a function of the PGU conversion efficiency, evaluated for the four different economic scenarios indicated in Table 2.*

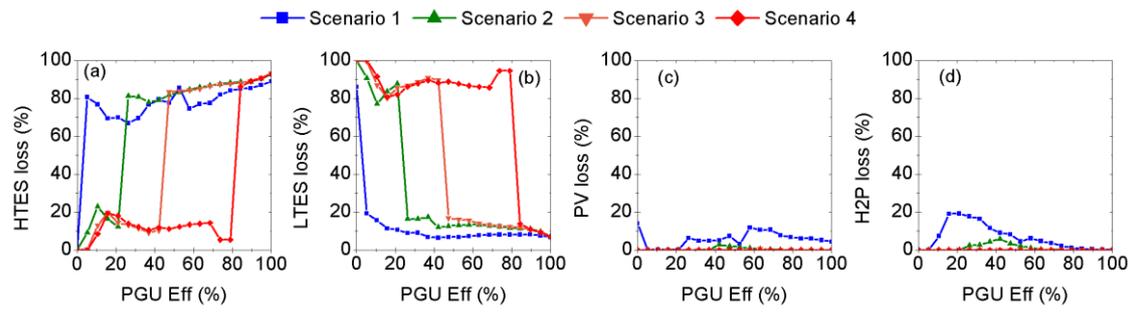

*Figure 5. Contribution of each kind of energy loss in a PHPS-E system optimized for each scenario indicated in Table 2: (a) Thermal insulation losses in the HTES, (b) thermal insulation losses in the LTES, (c) PV electricity that is not used nor stored, (d) waste heat from the PGU that is not stored in the LTES.*

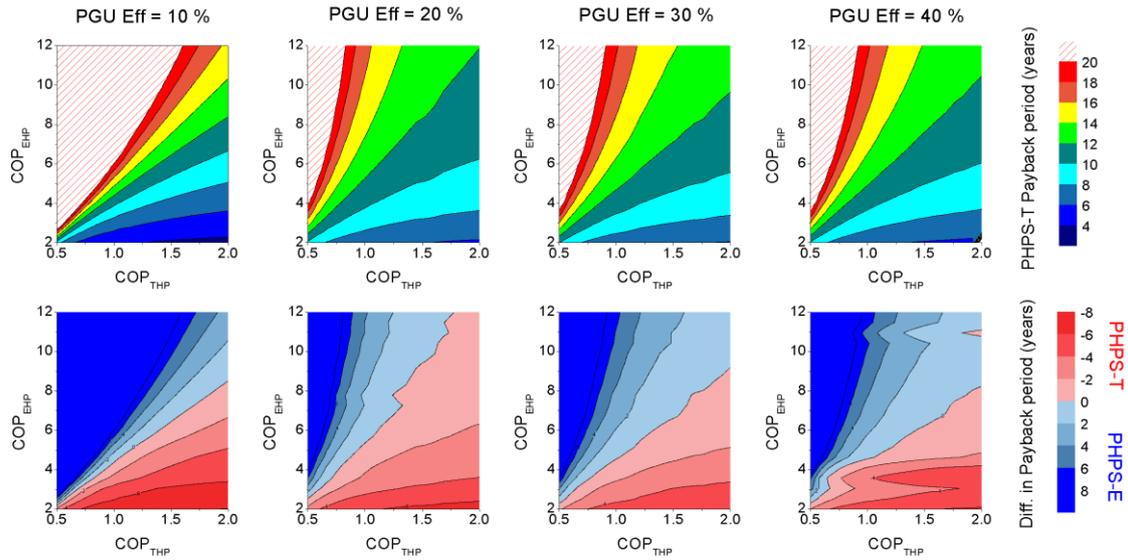

*Figure 6. (top) Contour plots representing the discounted payback period resulting from the installation of an optimized PHPS-T system as a function of the COP of THP (used in the PHPS-T system) and the COP of the EHP (used in the reference system); (bottom) contour plots representing the difference between the discounted payback period of a PHPS-E and a PHPS-T solution as a function of the COP of THP (used in the PHPS-T system) and the COP of the EHP (used in the PHPS-E system). Positive values (in blue) represent a PHPS-E solution with a shorter payback period than the PHPS-T one. Results are shown for four different PGU efficiencies. The rest of the parameters are taken from Table 1 and the Scenario 1 in Table 2.*

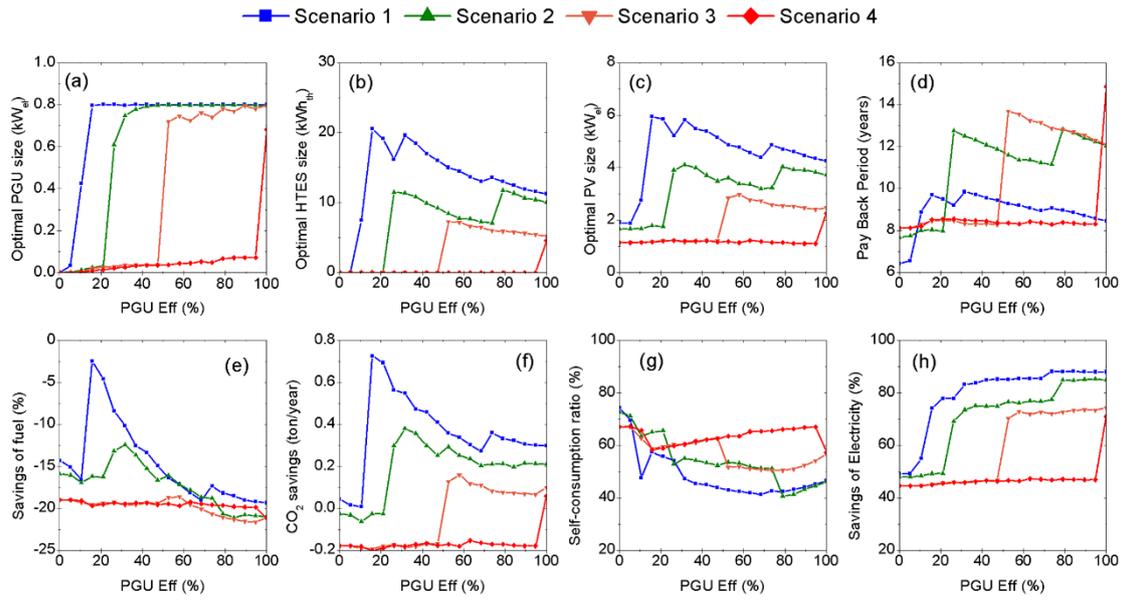

*Figure 7. Performance parameters of an optimized PHPS-T system (Figure 1-c) installed in Madrid as a function of the PGU conversion efficiency, evaluated for the four different economic scenarios indicated in Table 2.*

Table 1. Summary of the techno-economic variables used to describe the PHPS+PV system.

| Parameter | Value | Units |
|---|---|---|
| PGU conversion efficiency ($\eta_{PGU}$) | Independent variable | % |
| PV nominal installed power ($P_{nom,PV}$) | Variable (optimized) | $kW_{el}$ |
| HTES maximum capacity ($E_{HTES,max}$) | Variable (optimized) | $kWh_{th}$ |
| PGU maximum generation capacity ($P_{max,PGU}$) | Variable (optimized) | $kW_{el}$ |
| Inflation for electricity and fuel ($Infl_e = Infl_{el} = Infl_{fuel}$) | Variable (Table 2) | % |
| Weighted average cost of capital ($WACC_{nom}$) | Variable (Table 2) | % |
| CAPEX of HTES ($CAPEX^*_{HTES}$) | Variable (Table 2) | €/$kWh_{th}$ |
| CAPEX of PGU ($CAPEX^*_{PGU}$) | Variable (Table 2) | €/$kW_{el}$ |
| CAPEX of PV ($CAPEX^*_{PV}$) | Variable (Table 2) | €/$kW_{el}$ |
| CAPEX of LTES ($CAPEX^*_{LTES}$) | 30 [75], [76] | €/$kWh_{th}$ |
| CAPEX of EHP and THP ($CAPEX^*_{EHP}= CAPEX^*_{THP}$) | 500 [7][64] | €/$kW_{cool}$ |
| LTES temperature ($\Delta T_{LTES}$) | 70 (for PHPS-E) 130 (for PHPS-T) | ºC |
| HTES temperature ($\Delta T_{HTES}$) | 1200 | ºC |
| Parameter "$a_{LTES}$" for heat losses | 0.1 | $W \cdot K^{-1} dm^{-3/2}$ |
| Parameter "$a_{HTES}$" for heat losses | 0.1 | $W \cdot K^{-1} dm^{-3/2}$ |
| Overall inflation ($Infl$) | 2 | % |
| LTES maximum capacity ($E_{LTES,max}$) | 20 | $kWh_{th}$ |
| Installation lifetime ($T$) | 25 | years |
| $CO_2$ emissions of external fuel heating | 250 [61] | $gCO_2eq/kWh_{th}$ |
| $CO_2$ emissions of external grid electricity | 340 [63] | $gCO_2eq/kWh_{el}$ |
| $CO_2$ life-cycle equivalent emissions of PV electricity | 20 [60] | $gCO_2eq/kWh_{el}$ |
| $CO_2$ life-cycle equivalent emissions of HTES, LTES and PGU | Neglected [*] | $gCO_2eq/kWh_{el}$ |
| COP of EHP ($COP_{EHP}$) | 4 [7] | - |
| COP of THP ($COP_{THP}$) | 1.3 [35], [66], [67] | - |
| Variable cost of grid electricity ($OPEX^*_{elec,var}$) | 0.17 | €/$kWh_{el}$ |
| Fixed cost of grid electricity ($OPEX^*_{elec,fix}$) | 50 | €/$kW_{el}$-year |
| Variable cost of fuel ($OPEX^*_{fuel,var}$) | 0.07 | €/$kWh_{th}$ |
| Fixed cost of fuel ($OPEX_{fuel,fix}$) | 60 | €/year |

[*] Due to the difficulty on estimating the life cycle emissions for the HTES, LTES and PGU devices, we neglect their contribution. Thus, the $CO_2$ equivalent emissions will need to be corrected accordingly, when confident data become available.

Table 2. Economic scenarios considered in the study

| Scenario | $WACC_{nom}$ (%) | $Infl_e$ (%) | $CAPEX_{HTES}$ (€/kWh$_{th}$) | $CAPEX_{PGU}$ (€/kW$_{el}$) | $CAPEX_{PV}$ (€/kW$_{el}$) |
|---|---|---|---|---|---|
| Scenario 1 | 2 | 4 | 30 | 300 | 900 |
| Scenario 2 | | | 100 | 1000 | 1200 |
| Scenario 3 | 3 | 2 | | | |
| Scenario 4 | | | 200 | 2000 | |

Table 3. Summary of the results for the two kinds of systems (PHPS-E and PHPS-T) indicated in Figure 1, and the four scenarios described in Table 2. The rest of parameters are set to the values indicated in Table 1.

| Scenario | PGU Eff (%) | Optimized variables | | | Merit figures | | | | | |
|---|---|---|---|---|---|---|---|---|---|---|
| | | PGU size (kWel) | HTES size (kWhth) | PV size (kWel) | Total CAPEX [*] (k€) | Discounted payback period (years) | PV self-consumption ratio (%) | Electricity saves (%) | Fuel saves (%) | $CO_2$ emission saves (ton/year) |
| PHPS-E Scenario 1 | 60 | 3.11 | 49.02 | 11.48 | 12.74 | 11.82 | 41.0 | 95.3 | 19.0 | 1.69 |
| | 40 | 2.15 | 62.87 | 11.62 | 12.99 | 12.99 | 38.6 | 91.7 | 18.0 | 1.59 |
| | 20 | 1.25 | 29.47 | 8.37 | 8.79 | 12.22 | 48.7 | 69.3 | 19.8 | 1.46 |
| PHPS-E Scenario 2 | 60 | 0.94 | 16.67 | 6.20 | 10.05 | 14.14 | 50.2 | 75.6 | 9.6 | 1.20 |
| | 40 | 0.82 | 22.20 | 7.63 | 12.20 | 15.51 | 48.1 | 75.0 | 14.8 | 1.36 |
| | 20 | 0.16 | 3.85 | 2.29 | 3.28 | 14.70 | 49.7 | 28.1 | 3.7 | 0.45 |
| PHPS-E Scenario 3 | 60 | 0.79 | 8.53 | 3.47 | 5.81 | 17.11 | 56.2 | 54.3 | 4.4 | 0.80 |
| | 40 | 0.04 | 0.01 | 1.33 | 1.64 | 13.45 | 70.4 | 20.5 | 3.8 | 0.39 |
| | 20 | 0.02 | 0.01 | 1.45 | 1.76 | 13.54 | 70.4 | 20.7 | 4.5 | 0.41 |
| PHPS-E Scenario 4 | 60 | 0.04 | 0.01 | 1.34 | 1.68 | 13.53 | 71.6 | 20.9 | 3.8 | 0.39 |
| | 40 | 0.04 | 0.00 | 1.34 | 1.69 | 13.66 | 70.6 | 20.6 | 3.9 | 0.39 |
| | 20 | 0.02 | 0.01 | 1.40 | 1.71 | 13.84 | 69.0 | 20.2 | 4.1 | 0.39 |
| PHPS-T Scenario 1 | 60 | 0.80 | 14.18 | 4.68 | 4.88 | 9.14 | 42.3 | 85.6 | -17.6 | 0.32 |
| | 40 | 0.80 | 17.19 | 5.50 | 5.71 | 9.59 | 45.9 | 84.8 | -12.5 | 0.49 |
| | 20 | 0.80 | 19.37 | 6.02 | 6.24 | 9.52 | 56.6 | 77.9 | -3.2 | 0.74 |
| PHPS-T Scenario 2 | 60 | 0.80 | 7.73 | 3.36 | 5.60 | 11.38 | 52.6 | 76.4 | -17.5 | 0.24 |
| | 40 | 0.79 | 10.24 | 3.82 | 6.39 | 12.17 | 53.8 | 75.0 | -14.6 | 0.32 |
| | 20 | 0.04 | 0.01 | 1.75 | 2.14 | 7.95 | 66.4 | 49.3 | -16.2 | -0.02 |
| PHPS-T Scenario 3 | 60 | 0.71 | 6.82 | 2.85 | 4.81 | 13.36 | 51.3 | 71.8 | -19.2 | 0.13 |
| | 40 | 0.04 | 0.00 | 1.32 | 1.62 | 8.47 | 63.2 | 47.4 | -18.8 | -0.14 |
| | 20 | 0.02 | 0.01 | 1.23 | 1.50 | 8.51 | 59.7 | 45.8 | -19.3 | -0.18 |
| PHPS-T Scenario 4 | 60 | 0.05 | 0.00 | 1.14 | 1.46 | 8.33 | 63.9 | 46.5 | -19.7 | -0.18 |
| | 40 | 0.04 | 0.00 | 1.18 | 1.49 | 8.44 | 61.6 | 46.2 | -19.5 | -0.18 |
| | 20 | 0.02 | 0.01 | 1.19 | 1.46 | 8.55 | 58.5 | 45.4 | -19.6 | -0.19 |
| [*] **The addition of the CAPEX of PGU, HTES and PV systems** | | | | | | | | | | |

Table 4. Identical results than in Table 3, but for the case of loss-less (adiabatic) HTES and LTES (i.e. $a_{HTES} = a_{LTES} = 0$). The values in this table represent the upper bound for performance (unattainable in practice) of PHPS concept under the selected scenarios.

| Scenario | PGU Eff (%) | Optimized variables | | | Merit figures | | | | | |
|---|---|---|---|---|---|---|---|---|---|---|
| | | PGU size (kWel) | HTES size (kWhth) | PV size (kWel) | Total CAPEX[*] (k€) | Discounted payback period (years) | PV self-consumption ratio (%) | Electricity saves (%) | Fuel saves (%) | $CO_2$ emission saves (ton/year) |
| PHPS-E Scenario 1 | 60 | 3.35 | 51.92 | 6.77 | 8.65 | 7.78 | 81.3 | 99.0 | 24.8 | 2.09 |
| | 40 | 2.78 | 52.84 | 8.67 | 10.22 | 8.80 | 70.4 | 97.5 | 30.7 | 2.25 |
| | 20 | 1.32 | 32.09 | 7.31 | 7.93 | 9.26 | 73.5 | 78.5 | 28.8 | 1.96 |
| PHPS-E Scenario 2 | 60 | 2.13 | 31.25 | 5.93 | 12.37 | 11.69 | 85.9 | 95.4 | 22.1 | 1.96 |
| | 40 | 0.87 | 20.47 | 5.86 | 9.95 | 11.25 | 81.6 | 81.5 | 22.7 | 1.80 |
| | 20 | 0.71 | 20.14 | 5.20 | 8.97 | 12.65 | 80.9 | 63.3 | 22.1 | 1.56 |
| PHPS-E Scenario 3 | 60 | 0.78 | 13.74 | 4.05 | 7.01 | 11.27 | 93.4 | 78.4 | 14.4 | 1.49 |
| | 40 | 0.78 | 13.53 | 3.77 | 6.66 | 12.69 | 92.5 | 68.7 | 14.2 | 1.36 |
| | 20 | 0.17 | 6.76 | 2.01 | 3.25 | 12.81 | 98.8 | 32.4 | 9.8 | 0.75 |
| PHPS-E Scenario 4 | 60 | 0.67 | 12.24 | 3.65 | 8.16 | 14.28 | 95.2 | 73.0 | 13.1 | 1.38 |
| | 40 | 0.30 | 6.29 | 2.32 | 4.64 | 14.16 | 99.0 | 44.1 | 9.7 | 0.90 |
| | 20 | 0.03 | 0.27 | 1.42 | 1.81 | 11.55 | 100 | 21.1 | 7.5 | 0.53 |
| PHPS-T Scenario 1 | 60 | 0.80 | 17.69 | 3.11 | 3.57 | 5.86 | 100 | 93.3 | -10.7 | 0.74 |
| | 40 | 0.80 | 19.62 | 3.94 | 4.38 | 6.25 | 100 | 91.8 | -2.8 | 1.00 |
| | 20 | 0.80 | 24.01 | 5.34 | 5.77 | 6.98 | 97.0 | 86.6 | 9.6 | 1.38 |
| PHPS-T Scenario 2 | 60 | 0.79 | 9.55 | 2.94 | 5.27 | 8.39 | 100 | 90.4 | -11.6 | 0.67 |
| | 40 | 0.80 | 12.62 | 3.42 | 6.16 | 9.06 | 100 | 87.7 | -6.5 | 0.82 |
| | 20 | 0.47 | 14.78 | 3.68 | 6.36 | 9.67 | 99.7 | 74.4 | -0.9 | 0.85 |
| PHPS-T Scenario 3 | 60 | 0.79 | 8.83 | 2.48 | 4.64 | 9.51 | 100 | 86.8 | -14.8 | 0.51 |
| | 40 | 0.72 | 11.13 | 2.84 | 5.24 | 10.30 | 100 | 82.4 | -10.4 | 0.61 |
| | 20 | 0.13 | 5.56 | 1.54 | 2.53 | 9.12 | 100 | 54.3 | -15.1 | 0.09 |
| PHPS-T Scenario 4 | 60 | 0.61 | 6.98 | 2.18 | 5.23 | 11.78 | 100 | 80.6 | -15.9 | 0.39 |
| | 40 | 0.20 | 4.77 | 1.69 | 3.39 | 10.44 | 100 | 61.4 | -15.5 | 0.17 |
| | 20 | 0.02 | 0.18 | 1.17 | 1.49 | 7.48 | 100 | 45.8 | -16.3 | -0.06 |

[*] **The addition of the CAPEX of PGU, HTES and PV systems**